\documentclass[aps,pra,reprint,longbibliography,superscriptaddress,notitlepage]{revtex4-1}

\usepackage{amsmath,amsthm,amsfonts,amssymb,amscd}
\usepackage{fullpage}
\usepackage{lastpage}
\usepackage{graphicx}
\usepackage{hyperref}
\usepackage{wrapfig}
\usepackage{enumerate}
\usepackage{fancyhdr}
\usepackage{mathrsfs}
\usepackage{mathtools}
\usepackage{braket}
\usepackage{slashed}
\usepackage{bm}
\usepackage{float}
\usepackage[margin=0.55in]{geometry}
\usepackage[english]{babel}

\graphicspath{{../figs/}}

\begin{document}

\title{Geometry of epithelial cells provides a robust method for image based inference of stress within tissues.}
\author{Nicholas Noll}
\affiliation{Department of Physics, University of California Santa Barbara}
\affiliation{Biozentrum, University of Basel}
\affiliation{Swiss Institute of Bioinformatics}
\author{Sebastian J. Streichan}
\affiliation{Department of Physics, University of California Santa Barbara}
\affiliation{Kavli Institute for Theoretical Physics}
\author{Boris I. Shraiman}
\affiliation{Department of Physics, University of California Santa Barbara}
\affiliation{Kavli Institute for Theoretical Physics}

\begin{abstract}
Cellular mechanics plays an important role in epithelial morphogenesis, a process wherein cells reshape and rearrange to produce tissue-scale deformations. However, the study of tissue-scale mechanics is impaired by the difficulty of direct measurement of stress {\it in-vivo}. Alternative, image-based inference schemes aim to estimate stress from snapshots of cellular geometry but are challenged by sensitivity to fluctuations and measurement noise as well as the dependence on boundary conditions. Here we overcome these difficulties by introducing a new variational approach - the Geometrical Variation Method (GVM) - which exploits the fundamental duality between stress and cellular geometry that exists in the state of mechanical equilibrium of discrete mechanical networks that approximate cellular tissues. In the Geometrical Variation Method, the two dimensional apical geometry of an epithelial tissue is approximated by a 2D tiling with Circular Arc Polygons (CAP) in which the arcs represent intercellular interfaces defined by the balance of local line tension and pressure differentials between adjacent cells. We take advantage of local constraints that mechanical equilibrium imposes on CAP geometry to define a variational procedure that extracts the best fitting equilibrium configuration from images of epithelial monolayers. The GVM-based stress inference algorithm has been validated by the comparison of the predicted cellular and mesoscopic scale stress and measured myosin II patterns in the epithelial tissue during early \emph{Drosophila} embryogenesis. GVM prediction of mesoscopic stress tensor correlates at the 80\% level with the measured myosin distribution and reveals that most of the myosin II activity is involved in a static internal force balance within the epithelial layer. In addition to insight into cell mechanics, this study provides a practical method for non-destructive estimation of stress in live epithelial tissues. 
\end{abstract}

\maketitle
Cell and tissue mechanics is an important factor that both affects and regulates animal and plant development and thus is a subject of active study in developmental biology and biophysics, reviewed extensively in \cite{Far07,Rauz08,Blanchard09,Etournay16,Ingber05,Shraiman05,Pan16}. Here we focus on the mechanics of animal epithelial cells that compose tissues in the form of two-dimensional monolayers with tight junctions between adjacent cells and adhesion (of the basal cellular surface) to the substrate extracellular matrix (ECM) \cite{Frantz2010}. 

In the absence of a rigid substrate, mechanical properties of such monolayers are dominated by the tissue-wide mechanical network formed by cytoskeletal cortices coupled by intercellular adherens junctions \cite{AJ_review,lodish2008molecular,Halbleib06,Frank05}. Cytoskeletal cortices are localized to the lateral sides of cells, just below the apical surface, and are made of actin fibers cross-linked by myosin II motors that actively generate tension within the cortex.
The shape of cells within the tissue is determined by the balance of local actomyosin cytoskeletal contractility and the intracellular osmotic pressure \cite{Rauz08,AJ_review}, which acts to oppose the decrease in total cellular volume\cite{Paluch2012, Hyman2011}. For the purpose of tissue-scale mechanics, the full three-dimensional force balance that shapes individual cells can be approximated by an effective two-dimensional model of the apical cytoskeletal network. In this simplified 2D view, the contractility of the junctional actomyosin ``belts" balances against an effective two-dimensional pressure that prevents the collapse of the apical area under cortical tension: this type of an approximation underlies the widely used ``vertex model" approach to epithelial cell mechanics \cite{Honda83, Huf07, Far07, Noll17}.

Measuring mechanical properties of cells and tissues in-vivo presents a considerable experimental challenge. AFM \cite{AFM_review} and optical tweezer contact microscopy \cite{Bam09} have been used to probe the local rheology of individual cellular interfaces at great resolution but do not provide a direct readout of internal stress. 
The most common method for detecting  stress {\it in vivo} is UV laser ablation, in which focused light `cuts' the cytoskeletal bundle abutting a cell-cell interface and the resultant retraction velocity is used as a proxy for the local cortical tension \cite{Bonnet12}. This method is convenient, as it does not require any special preparation of the sample, but is destructive and hence does not allow measurement of the global stress distribution across the tissue. Other methods use genetically encoded FRET tension sensors engineered into load carrying proteins \cite{FRET_Grashoff} or employ measurements of deformation with implanted beads\cite{Campas14,Doub17}. These methods require specially prepared samples and are technically challenging, both in implementation and in quantitative interpretation.

The difficulty of direct experimental measurement of mechanical stress in developing tissues has stimulated alternative approaches that seek to capitalize on the availability of live imaging data \cite{Brodland10,Chiou12,Ishihara12}. For example \cite{Chiou12,Ishihara12} introduced a method for inferring cellular stress from observed 2D cell geometry based on the assumption that the tissue is instantaneously in a mechanical equilibrium, described by  a model parameterized directly by intercellular pressure and tension. In the simplest versions of the method \cite{Chiou12,Ishihara12} 2D geometry was parametrized by a polygonal tiling generally used in Vertex Models \cite{Honda83,Far07}. The validity of this approach, of course, rests upon the accuracy of the assumptions and approximations, which varies between tissues and conditions and has to be evaluated in each case. The major challenge to the approach stems from the sensitivity to both noise in the measurement of cellular geometry and the prior distribution imposed on the boundary conditions at the edge of the observed tissue domain. These difficulties have necessitated additional assumptions, introduced to over-constrain the inference problem \cite{Chiou12,Ishihara12}. A particularly direct generalization aims to extract additional information from the image data, most obviously provided by the observable curvature of cell interfaces \cite{Chiou12,CellFit}: an example of such an approach is the CellFIT toolkit \cite{CellFit}. Sensitivity to noise and image quality however remains a major issue. Below, we develop a new approach, improving the image-based ``mechanical inference"  to the extent that makes it broadly applicable in the practice of experimental data analysis.

In this paper, we explore general constraints imposed by mechanical equilibrium on the geometry of 2D cellular arrays that balance arbitrary interfacial tension against differential cellular pressure. Our analysis will identify a certain mathematical duality between the geometry of cells and a triangulation formed by the equilibrium values of interfacial tensions. This duality provides a set of highly nontrivial constraints that can be used to both effectively de-noise image analysis and stabilize the problem of mechanical inference, which is achieved through a variational formulation of the inference problem.
Using synthetic data as a comparative benchmark, we show that our algorithm correctly infers mechanics under arbitrary pressure differentials and moderate measurement noise, performing significantly better than existing methods. To illustrate its practical utility, we apply the algorithm to live imaging data from the early stages of \emph{Drosophila} embryonic development and demonstrate its ability to accurately predict - based on cell geometry alone - the spatial distribution and anisotropy of myosin II, the molecular motor known as a generator of mechanical stress in the developing embryo \cite{Streichan17,Bing14,Collinet15,Rauz08}.
Synthetic and real data tests suggest practical utility for the new mechanical inference algorithm as a tool for quantifying stress distribution in live tissue in the absence of direct measurement of local forces.

\subsection*{Equilibrium properties of 2D cell arrays with non-uniform pressure}

\begin{figure}[ht]
\centerline{
\includegraphics[width=.5\textwidth]{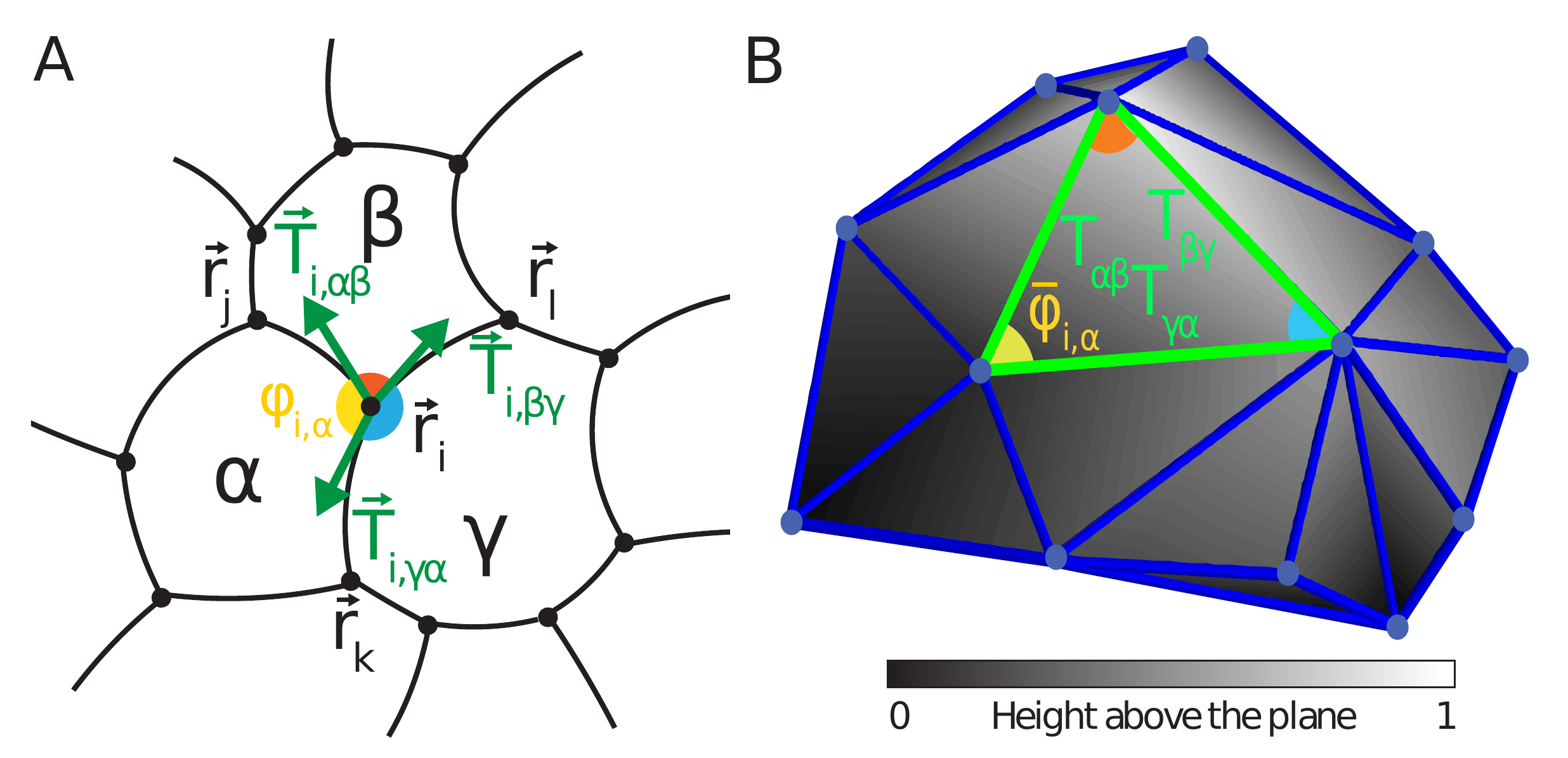}
}
\caption{\textbf{Circular Arc Polygonal (CAP) tiling and its dual tension triangulation.} A) Circular arc polygons provide an approximate representation of  equilibrium geometry of a cell array.  Curvature of interfaces, is controlled by pressure differences between adjacent cells and is related to interfacial tension by the Young-Laplace law \cite{landau}. B) Force balance at vertex $i$ requires the three tension vectors tangent to each edge to sum up to zero, which defines a local triangle. As adjacent vertices in A) share edges, the tension triangles of each vertex form a triangulated surface - the dual representation of force balance - with each triangular face corresponding to each vertex. In the absence of pressure differentials the dual surface would be flat: height variation arises from differential pressure.}
\label{capProperties}
\end{figure}

In order to model the mechanical state of the 2D epithelial tissue, we generalize the standard Vertex Model \cite{Honda83} which represents the epithelium by a planar polygonal tiling parameterized solely by the positions of vertices, the location where three or more cells meet, hereafter denoted $\bm{r}_i$. Here, we shall approximate the geometry of cells in the epithelial layer by a Circular Arc Polygonal (CAP) tiling, replacing the straight polygonal edges with circular arcs that correspond to tensed interfaces balancing pressure differentials between the adjacent cells as described by the Young-Laplace (Y-L) law \cite{landau}.  The equilibrium geometry of a CAP tiling is fully specified by the set of effective interfacial tensions ($T_{\alpha\beta}$, where $\alpha,\beta$ labels the cells partitioned by the given edge) and the set of effective hydrostatic pressures $p_\alpha$ representing the contribution of bulk stress from the cell to the 2D apical force balance. In mechanical equilibrium the radius of the circular arc forming $\alpha \beta$ interface, $R_{\alpha\beta}$, is given by the Y-L Law, $R_{\alpha\beta} = T_{\alpha\beta}/[p_{\alpha} - p_{\beta}]$. A graphical example of the resultant CAP network is shown in Fig. \ref{capProperties}. 

The dimensionality of the space of all CAP tilings is given by the number of internal degrees of freedom: two positional degrees of freedom for each vertex plus the radius of curvature for each edge. The total count is $2v+e=7c$ where $c,e,v$ respectively denote the number of cells, edges and vertices within the tiling, which satisfy $v=2c$ and $e=3c$. (This count is exact on a torus and is approximate up to the boundary corrections, for a large planar array).  Conversely, the mechanical state of the network is parameterized by $e+c=4c$ parameters corresponding to independent interfacial tensions and cell pressures. Hence,  the dimension of the space of all \emph{equilibrium} CAP tilings is $4c$ and thus, mechanical equilibrium implies, on average, three constraints per cell. Our analysis below will aim to i) define a relation between the tiling geometry and tensions and pressures and ii) uncover the geometric constraints associated with mechanical balance. 

Mechanical equilibrium of a CAP network is reached when tensions balance at every vertex and interfacial curvature obeys the Young-Laplace Law. Tension acts tangentially along network edges, Fig. \ref{capProperties} depicts an example of force balance for vertex $i$ due to tensions of interfaces that connect adjacent vertices $j,k,l$.  We note that an edge can be labeled in two equivalent ways: either with the labels of cells the edge partitions or the vertex labels that the edge connects, e.g. edge $ij$ in Fig. \ref{capProperties}a separates cells $\alpha,\beta$, so we shall define ${T}_{ij}={T}_{\alpha\beta}$ and use both schemes interchangeably. Hence force balance requires
\begin{equation} \label{FB}
\frac{T_{ij}}{R_{ij}} \! \left[ \bm{r}_i - \bm{\rho}_{ij} \right]^* \! + \! \frac{T_{ik}}{R_{ik}} \!\left[ \bm{r}_i - \bm{\rho}_{ik} \right]^* \! + \! \frac{T_{il}}{R_{il}}\! \left[ \bm{r}_i - \bm{\rho}_{il} \right]^*  = 0
\end{equation}
where $\bm{\rho}_{ij}$ and $R_{ij}$ is defined as the centroid and signed curvature of the circular arc associated to edge connecting vertices $i,j$.  The *-superscript here is shorthand notation for the counter-clockwise rotation by $\pi/2$, e.g. $\bm{r}^* \equiv \bm{\hat{z}} \wedge \bm{r}$. Each of the three terms is recognized as a tension vector, e.g. ${\bm T}_{ij}=T_{ij}(\bm{r}_i - \bm{\rho}_{ij})^*/{R_{ij}} $, acting on vertex $i$. Hence, force balance can be re-interpreted geometrically as a  triangle formed by tension vectors acting on vertex $i$, with adjacent cells $\alpha, \beta, \gamma$, see Fig. \ref{capProperties} being associated with the vertices of the tension triangle. Crucially, as local force balance triangles associated with adjacent vertices share an edge, stitching together force balance conditions on all network vertices defines a tension triangulation `dual' to the CAP tiling. Dual triangulation vertices correspond to cells and triangular faces correspond to the vertices of the original CAP tiling, as shown in Fig. \ref{capProperties}.

The tension triangulation dual to a CAP tiling is not planar. From the definition of the dual triangulation, the triangle angles are simply related to the angles at CAP vertices: ${\bar \phi}_{i\alpha} = \pi-{ \phi}_{i\alpha}$. If the edges of the tiling polygons were straight, one would find that $\sum_i {\bar \phi}_{i\alpha} =2\pi$, corresponding to a planar triangulation\cite{Noll17}.  
As edges are curved, tension vectors acting at either end of an interface are no longer parallel: 
 the force vector acting on vertex $i$ due to interface $\alpha \beta$ is rotated by an angle $\Delta \varphi_{\alpha\beta} = \ell_{\alpha\beta}/R_{\alpha\beta}$ from the equivalent force acting on vertex $j$, where $\ell_{\alpha\beta}$ is the physical arc length of the circular edge. This results in the non-planarity of the tension-triangulation manifested by the ``deficit" angle  for each cell, $\Delta \varphi_\alpha$,  defined as the sum of curvature contributions from all edges that compose the cell:
\begin{equation} \label{defAngle}
\Delta \varphi_\alpha=\sum_{\{\beta \}_\alpha} \frac{\ell_{\alpha \beta}}{R_{\alpha \beta}}
\end{equation} 
The deficit angle $\Delta \varphi_\alpha$ associated with cell $\alpha$ is the discrete Gauss curvature for the corresponding vertex in the dual triangulation.

We now proceed to define equilibrium constraints on CAP geometry and explicit relations between tensions and pressures and geometric observables. By applying the Sine Law to a triangle  one can relate the ratio of tensions in adjacent edges to the corresponding angles, e.g. for the triangle $i$: $T_{\alpha\beta}/T_{\alpha\gamma}  =  \sin(\varphi_{i,\gamma})/\sin(\varphi_{i,\beta})$. Multiplying such ratios for a set of triangles that share a vertex of the dual triangulation uncovers a non-trivial constraint on CAP tiling angles
\begin{equation} \label{genCompCond}
\chi_\alpha \equiv \displaystyle\prod_{i \in V_\alpha} \frac{\sin \varphi_{i,\gamma}}{\sin \varphi_{i,\beta}}= \displaystyle\prod_{i \in V_\alpha} \frac{T_{\alpha\gamma}}{T_{\alpha\beta}} = 1
\end{equation}
(where the product is taken over the set $\mathcal{V}_\alpha$ of vertices $i$ that belong to cell $\alpha$, while $\beta$ and $\gamma$ label other cells adjacent to $i$ in clockwise order). Eq. (\ref{genCompCond}) defines $c$ non-trivial constraints on the angles of an equilibrium CAP network which we recognize as the generalized form of the geometric compatibility condition introduced in \cite{Noll17}. Provided that the geometric constraints on the CAP tiling given by Eq. (\ref{genCompCond}) are satisfied, the dual tension triangulation specifies all $T_{\alpha\beta}$ up to an overall scale - the key property making inference possible. 

Next, given the set of tensions $T_{\alpha \beta}$ from the tension-triangulation, pressures can be computed on the basis of the Young-Laplace Law by solving a discrete Poisson equation on the dual triangulation:
\begin{equation} \label{pressurePoisson}
\sum_{\{\beta \}_\alpha} (p_\alpha -p_{ \beta})=\sum_{\{\beta \}_\alpha} \frac{T_{\alpha \beta}}{R_{\alpha \beta}}
\end{equation} 
Eq. (\ref{pressurePoisson}) represents $c$ equations that define $c$ pressure unknowns up to the homogeneous solution which has to be fixed by the boundary conditions on $p_\alpha$. 

Importantly, the Y-L law  implies that $T_{\alpha \beta} / R_{\alpha \beta} + T_{\beta \gamma} / R_{ \beta \gamma}  + T_{\gamma \alpha } / R_{\gamma \alpha } = 0$ must be satisfied at each vertex (e.g. vertex $i$ shared by cells $\alpha, \beta, \gamma$). Using the Sine Law on the dual triangulation, this constraint is recast in the purely geometric form
\begin{equation} \label{YLconstraint}
\frac{\sin(\varphi_{i,\gamma})}{R_{\alpha \beta}  }+ \frac{\sin(\varphi_{i,\alpha})}{R_{ \beta \gamma}  }+ \frac{\sin(\varphi_{i,\beta})}{R_{\gamma \alpha }}=0
\end{equation} 
which defines $v=2c$ (one per each vertex) geometric constraints that account for the difference in the number of pressure variables ($c$) and the number of $R_{\alpha\beta}$ variables ($e=3c$). Together, Eqs. (\ref{genCompCond}, \ref{YLconstraint}) impose $3c$ constraints on a equilibrium CAP network  so that the dimensionality of the latter is indeed given by $4c$, which is equal to the total number of tension and pressure variables.

In principle, one can infer the underlying balanced mechanical state - the set of equilibrium interfacial tensions $\{T_{\alpha\beta}\}$ and pressures $\{p_\alpha\}$ - by building the tension-triangulation: utilizing local network angles at each vertex to build each dual triangular face, and subsequently solving Eq. (\ref{pressurePoisson}) to obtain cell pressures. This approach is contingent upon the network satisfying compatibility conditions Eqs. (\ref{genCompCond},\ref{YLconstraint}). In practice it suffers from two major problems: (i) real cellular networks undergoing morphogenesis are expected to be close to, but not exactly in, mechanical equilibrium, (ii) the measurement of network geometry from imaging data will always be noisy and imperfect. As a result, an algorithm attempting to stitch together a \emph{global} tension triangulation would rapidly accumulate errors that would dramatically impair the resultant inference. Furthermore, the solution of Eq. (\ref{pressurePoisson}) requires the knowledge of boundary conditions which are generally not known. 

\subsection*{Unconstrained parameterization of equilibrium CAP networks}

\begin{figure}[ht] 
\centerline{
\includegraphics[width=.35\textwidth]{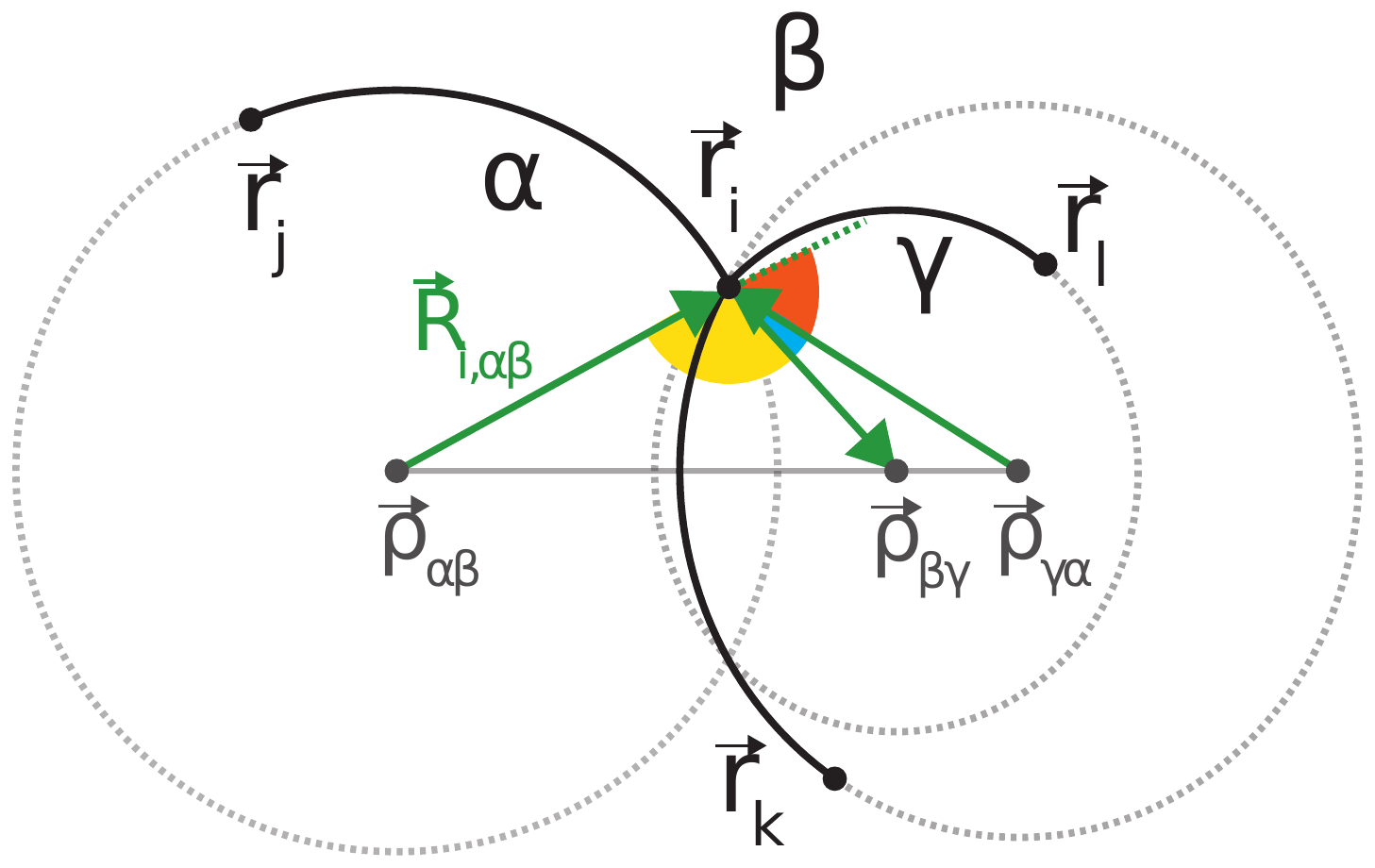}
}
\caption{\textbf{Parameterization of equilibrium CAP network.} The equilibrium condition at a vertex in a force-balanced cell array is equivalent to the collinearity of arc centroids.}
\label{forwardDef}
\end{figure}

To achieve a robust implementation of the mechanical inference we shall recast it as a variational problem on an unconstrained set of variables. The idea is to fit the observed network geometry to the closest equilibrium CAP tiling in the least-squares sense by utilizing a set of variables that both natively satisfies the constraints of equilibrium and fully characterizes the mechanical state of the tissue. 
We begin by parameterizing the CAP tiling by circular arc centroids $\bm{\rho}_{\alpha\beta}$ and the associated radii of curvature $R_{\alpha\beta}$ and minimize the variational pseudo-energy function
\begin{equation} \label{inverseEnergy}
E = \frac{1}{2 n_e}\!\displaystyle\sum\limits_{(\alpha,\!\beta)} \sum\limits_{n}^{N_{\alpha\beta}} {\left( \big|\bm{r}_{\alpha\beta}(n)\! - \!\bm{\rho}_{\alpha\beta} \big| \! - \! R_{\alpha\beta}\right)^2}
\end{equation}
$\bm{r}_{\alpha\beta}(n) $ denotes the position of $n^{th}$ pixel on edge $\alpha,\beta$ obtained directly from the segmentation and $N_{\alpha\beta}$ denotes the number of pixels segmented per edge. Eq. (\ref{inverseEnergy}) has a simple geometric interpretation - it penalizes the Euclidean distance between the estimated and measured circular arc for each pixel of a segmented edge. 

The $9c$ fitting variables in Eq. (\ref{inverseEnergy}) correspond to an arbitrary CAP tiling -- we want to restrict to the subset of equilibrium CAP tilings. To define the reduced set of degrees of freedom we revisit Eq. (\ref{FB}) and note that the prefactor for each tension vector can be reinterpreted using the Y-L Law, leading to a relation between the arc centroids of the three edges connected at each vertex 
\begin{equation} \label{centroidCollinear}
\left[p_\beta - p_\alpha\right] \bm{\rho}_{\beta\alpha} + \left[p_\alpha - p_\gamma\right] \bm{\rho}_{\alpha\gamma} + \left[p_\gamma - p_\beta\right] \bm{\rho}_{\gamma\beta} = 0
\end{equation}
Remarkably, at equilibrium, the centroids of all three interfacial circular arcs meeting at a vertex are constrained to be \emph{collinear}, with distance along the line controlled by relative pressure differences, depicted graphically in Fig. \ref{forwardDef}a.  This is a practically useful observation that provides a direct method to assess the compatibility of any cellular network with mechanical equilibrium solely on the basis of image analysis of cell morphology. 

Additionally, in order for all three edges, corresponding to collinear centroids $\bm{\rho}_{\alpha\beta}$, to intersect at a single point  defining a CAP vertex, their respective radii must obey 
\begin{align} \label{vertexDef}
&\left(p_\alpha - p_\beta\right) R_{\alpha\beta}^2 + \left(p_\beta - p_\gamma\right) R_{\beta\gamma}^2 + \left(p_\gamma - p_\alpha\right) R_{\gamma\alpha}^2  = \nonumber \\
&\left(p_\alpha - p_\beta\right) \bm{\rho}_{\alpha\beta}^2 + \left(p_\beta - p_\gamma\right) \bm{\rho}_{\beta\gamma}^2  + \left(p_\gamma - p_\alpha\right) \bm{\rho}_{\gamma\alpha}^2 
\end{align}
which imposes $v$ constraints on $e$ curvature variables, $R_{\alpha\beta}$, reducing the set to a $e-v=c$ independent degrees of freedom (as expected from the fact that edge curvatures are generated by intracellular pressures which make $c$ variables). Given any solution, we can generate another geometrically compatible cell array by transforming $\bar{R}_{\alpha\beta}^2 = R_{\alpha\beta}^2 + \frac{p_\alpha z^2_\alpha - p_\beta z^2_\beta}{p_\alpha-p_\beta}$, where $z_\alpha$ provide the explicit degrees of freedom for the $c$ dimensional manifold of solutions of Eq. (\ref{vertexDef}) that share the same set of edge centroids. 

We next observe that Eq. (\ref{centroidCollinear}) admits the general solution
\begin{equation} \label{centroidDecomp}
\bm{\rho}_{\alpha\beta} = \frac{p_\beta \bm{q}_\beta - p_\alpha \bm{q}_\alpha}{p_\beta - p_\alpha}
\end{equation}
where we have introduced a vector variable $\bm{q}_\alpha$ for each cell.  Substituting this form into Eq. (\ref{vertexDef}) we obtain an explicit general expression for $R_{\alpha\beta}$:
\begin{equation} \label{tensionDef}
    R_{\alpha\beta} \!=\! \sqrt{\frac{p_{\alpha}p_{\beta} |\bm{q}_\alpha\! -\! \bm{q}_\beta|^2 }{(p_\alpha\! - \!p_\beta)^2}\!-\! \frac{p_\alpha z_\alpha^2 \!-\! p_\beta z^2_\beta}{p_\alpha\! - \!p_\beta} } 
\end{equation}
Together Eqs. (\ref{centroidDecomp},\ref{tensionDef}) provide an explicit parameterization of equilibrium CAP tiling by an arbitrary choice of $4c$ independent variables $\{\bm{q}_\alpha,z_\alpha,p_\alpha\}$. These variables explicitly satisfy all geometric constraints and provide an explicit local expression for the tensions via the Y-L law. We note also that $\{\bm{q}_\alpha , z_\alpha \}$ variables have a nice geometric interpretation as the set of centroid points in 3D which generate equilibrium CAP tilings via generalized Voronoi construction defined in Appendix A, where we also derive an explicit relation between $\{\bm{q}_\alpha , z_\alpha \}$ and CAP vertices given by $p_\alpha(|\bm{r}_{i} - \bm{q}_\alpha|^2 + z^2_\alpha) =p_\beta(|\bm{r}_{i} - \bm{q}_\beta|+z^2_\beta)=p_\gamma(|\bm{r}_{i} - \bm{q}_\gamma|+z^2_\gamma)$, which can be solved to define $\bm{r}_{i}$ from the reduced variables $\{\bm{q}_\alpha,z_\alpha,p_\alpha\}$. Note that pressures (and hence tensions) are defined only up to an overall scale factor, which leaves geometry invariant.

\subsection*{The Geometrical Variation Method for inferring internal stress distribution.}

Eqns. (\ref{centroidDecomp}, \ref{tensionDef}) parameterize $\bm{\rho}_{\alpha\beta}$ and $R_{\alpha\beta}$ in terms of a reduced set of degrees of freedom, $\{\bm{q}_\alpha, p_\alpha, z_\alpha^2 \}$, that guarantee the geometric constraints of an equilibrium CAP network. Given a grayscale image of cellular boundaries in an epithelial monolayer, comprising a large number of cells (20-1000) with interfaces resolved at a pixel level, we can find the best approximation of the 2D network formed by cell interfaces. This is achieved by minimization of Eq. (\ref{inverseEnergy}) with respect to this reduced set of degrees of freedom produces the best fit of image data by an equilibrium CAP tiling.  Minimization of  Eq. (\ref{inverseEnergy}) is a non-linear optimization problem subject to linear inequalities on $z_\alpha^2$'s that ensure positivity of the argument of the square root in Eq. (\ref{tensionDef}). We solve the problem computationally using MATLAB's implementation of the interior-point algorithm.
The simple choice of starting the iteration with $z_\alpha=0$ and the estimate of $\{p_\alpha, \bm{q}_\alpha\}$ from the observed set of network angles (as explained in detail in the Appendix B) was found to produce reliable convergence. 

\begin{figure}[ht] 
\centerline{
\includegraphics[width=.5\textwidth]{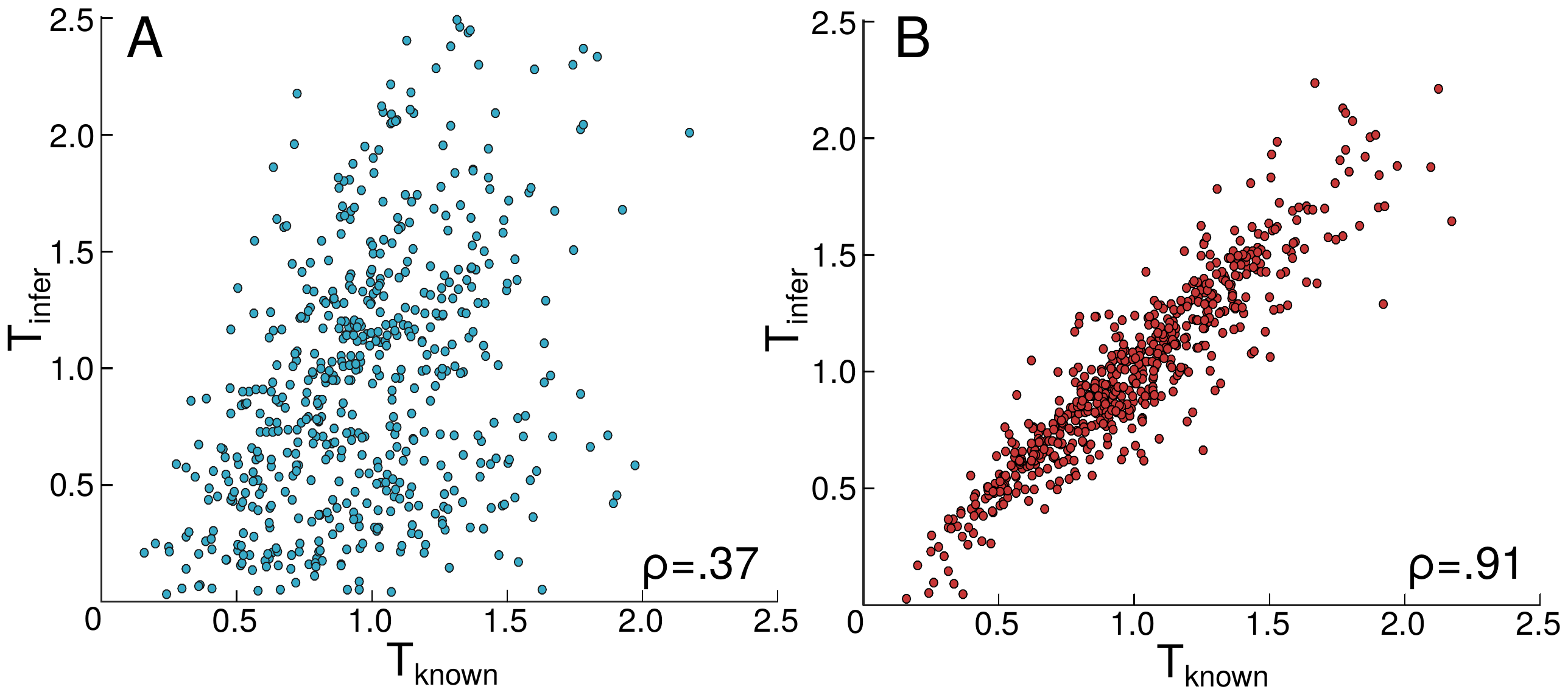}
}
\caption{\textbf{Validation of GVM for stress inference using synthetic data.} A) Comparison of actual and inferred tensions using the "matrix inverse" inference method introduced in \cite{Chiou12} made over-determined using measured edge curvature data, as in \cite{CellFit}. $10\%$ noise was added to the synthetic input data (see Appendix C for details). $\rho$ denotes the Pearson's correlation coefficient. B) Same as (A) but using the GVM on the same synthetic data. Note the significant increase of the correlation coefficient.}
\label{syntheticInverse}
\end{figure}

To evaluate the performance of GVM on mechanical inference, we tested it against synthetically generated data. As shown in Fig. \ref{syntheticInverse},  we find that inference of the dual triangulation is quite robust to both measurement noise and large inter-cellular pressure differentials (see the Appendix C for details). 
Key to the improved performance is the over-constrained nature of the present formulation of the mechanical inference problem which combines the estimation of geometric parameters with stress inference into a single variational analysis of image data. 
The redundancy does not only stabilize inference in the presence of noise, it also allows us to infer {\it boundary stresses} using just information of 2D bulk cellular morphology. The method is local and thus immediately applicable to inference of inter-tissue forces, as will be discussed below.

The inferred set of interfacial tensions and  cellular pressures  allows us to construct a stress tensor for an epithelial layer in mechanical equilibrium. Specifically, inside the bulk of cell $\alpha$, the stress tensor is isotropic and constant $\sigma^{ab} = p_\alpha \delta^{ab}$ (where $a,b$ index spatial dimensions) and $\sigma^{ab} = T_{\alpha\beta} \bm{\hat{r}}_{\alpha\beta}^a \bm{\hat{r}}_{\alpha\beta}^b \delta(|\bm{x}-\bm{\rho}_{\alpha\beta}| - R_{\alpha\beta})$ on edge $\alpha\beta$. One can average this quantity over cell  area $A_\alpha$:
\begin{equation} \label{stress}
\bar{\sigma}^{ab}_{\alpha} = -p_\alpha \delta^{ab} + \displaystyle\sum\limits_{\{\beta\}_\alpha} \frac{T_{\alpha\beta}}{2 A_\alpha} \displaystyle\int\limits_{\bm{r}_{\alpha\beta}} \!dr\, \bm{\hat{r}}_{\alpha\beta}^a \bm{\hat{r}}_{\alpha\beta}^b
\end{equation}
where $\{\beta\}_\alpha$ denotes all cells connected to cell $\alpha$ and integration is taken along the cell boundary arcs. In practice it is useful to coarse-grain the tensor by averaging over neighboring cells cells to obtain an approximation to the continuum stress tensor. In the following section, we verify the utility of the proposed mechanical inference by comparing the inferred stress tensor to known biological correlates of stress. 

\subsection*{Local and global stress inference and its {\it in vivo} correlates. }

We now apply GVM to imaging data from living epithelial monolayers. Ideally we would test the GVM inference against direct measurements of stress. Presently, the most reliable readout of local stress in live tissue is provided by observed levels of (fluorescent labelled) junctional myosin, which has been previously demonstrated to correlate with local interfacial tension measured by laser ablation \cite{Rauz08}. We shall carry out the comparison between inferred stress and observed myosin level first on the scale of cells, then on the scale of the whole tissue, using the data on \emph{Drosophila} embryonic development.

During the initial stages of \emph{Drosophila} embryonic development, the ellipsoidal monolayer of epithelial cells forming the embryo undergoes a series of non-trivial mechanical transformations. Immediately following the formation of the ventral furrow (VF) - the first step of gastrulation - \emph{Drosophila} embryo undergoes germ-band extension (GBE): a major morphogenetic movement involving a convergent extension of the lateral ectoderm,  which approximately doubles its length along the embryo's anterior-posterior (AP) axis. This process has been demonstrated to be driven by the activity of the junctional pool of myosin II, which exhibits a non-uniform and anisotropic distribution on the surface of the embryo, in particular, forming  contractile supracellular cables that run  along the dorso-ventral (DV) axis of the embryo \cite{Collinet15,Streichan17,Zallen08}. 
Laser ablation assays have demonstrated that these myosin cables, associated with DV oriented edges, exhibit significantly higher cortical tension than AP oriented cell junctions \cite{Rauz10,Zallen08}. 
The quantitative relation between myosin and mechanical stress was further elaborated in our earlier study of morphogentic flow  \cite{Streichan17} which demonstrated that a symmetric 2D tensor $m_{ab}$ describing coarse grained distribution of myosin is a useful proxy for the stress tensor.

Applying GVM to the embryonic epithelium images \cite{Streichan17} we found that cell-array geometry observed over the first 60 minutes of convergent extension, is quite well approximated by equilibrium cell network:  $\langle E \rangle \approx 1$, which means our best fit equilibrium CAP geometry differs from the image segmentation by on average one pixel per edge! 

\begin{figure*}[ht] 
\centerline{
\includegraphics[width=.8\textwidth]{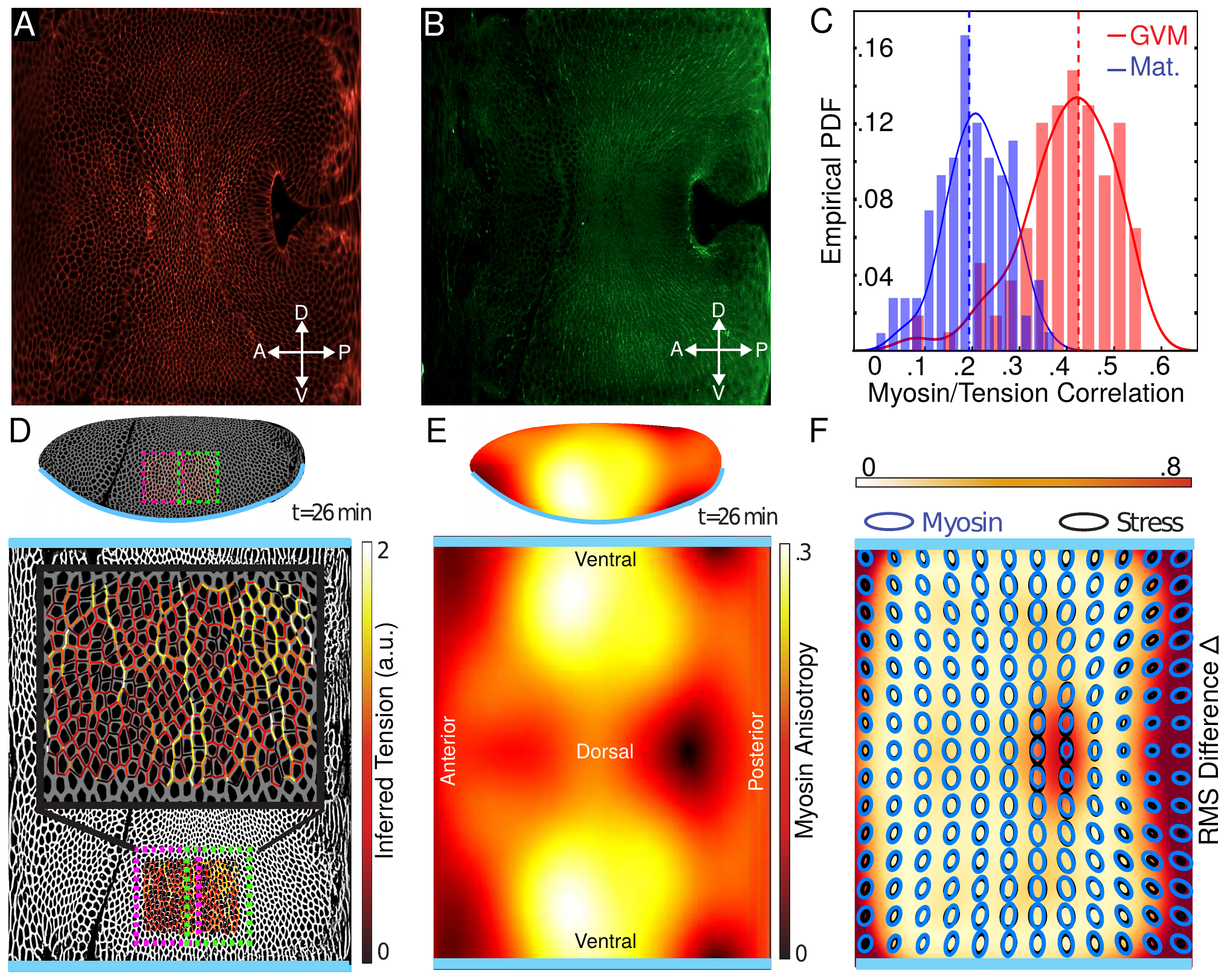}
}
\caption{\textbf{Stress inference results compared to measured myosin distribution for the germ-band extension in Drosophila embryo.}
A,B) Raw in-toto images showing fluorescent labelled membrane (mCherry, shown in A) and myosin (Sqh-GFP shown in B) obtained using light sheet microscopy and unrolled into the plane using ImSAnE \cite{Heemskerk15}. Time corresponds to roughly 26 minutes after cephalic furrow formation.
C) The correlation between myosin and inferred tensions comparing the matrix inverse method defined in \cite{Chiou12} and the new GVM-based inference algorithm during the first 40 minutes of GBE.
D) An example of stress inference, combining two overlapping image frames used for the analysis, shown directly on the embryo and on a cylindrical projection with the ventral line cut and mapped onto the top and bottom edges of the image with the dorsal side along the midline. 
Inset shows color-coded edge tension inferred by the GVM algorithm. The inferred stress displays stress cables as expected for the lateral region of the embryo at  the time considered.
E) The mesoscopic anisotropy of myosin, 26min post CF formation, 
shown on the embryo and its cylindrical projection.  Anisotropy is largest in the lateral region. The principal axis of myosin in this region points along the DV axis. 
F) 
The spatial distribution of the normalized r.m.s. difference ($\Delta(r)$) between the GVM inferred stress tensor and the measured total myosin tensor 
(at 26min post CF). Both tensor fields are represented as ellipses for direct comparison. The large discrepancy ($\Delta \sim .8$) found in the vicinity of the anterior and posterior poles (mapped to the left and right edges of the cylindrical projection) is due to the poor imaging of the poles. The large discrepancy in the center of the dorsal region is real and can be explained by the difference between the total and "balanced" myosin distributions as explained in the text. 
} 
\label{globalTiling}
\end{figure*}

Fig. \ref{globalTiling} shows results of the analysis of the lateral ectoderm during GBE. 
Qualitatively, inferred stress exhibits anisotropic stress cables that run along the DV axis in agreement with previous studies \cite{Zallen08}. 
For a quantitative comparison, we compute the correlation coefficient of tension inferred on individual cellular interfaces with the myosin line density measured on same interfaces: a histogram of the calculated results for each time point is shown in Fig.\ref{globalTiling}D. The mean correlation coefficient, $\rho \sim .4$, is a two-fold improvement over the earlier matrix inverse method \cite{Chiou12}.    

The observed correlation indicates that the inference method is picking up underlying mechanical effects. 
There are however numerous sources of noise that weaken correlation. To name two: i) while the analysis was carried out on a single snapshot, cell geometry is fluctuating on the timescale of seconds; ii) linear density of (fluorescent-labelled) myosin is not an exact measurement of line tension. 
Most importantly, the assumption that cells are in a mechanical equilibrium is at best only an approximation: in the case of GBE, there is a mean morphogenetic flow of cells indicating the presence of unbalanced local forces within the tissue\cite{Streichan17}. Below we shall demonstrate that accounting for this systematic deviation from mechanical equilibrium effect improves correlation between inferred stress and measured myosin distribution at the mesoscopic scale. 

Edge by edge comparison is the most exacting test as it is sensitive to local fluctuations. Myosin distribution however also exhibits non-trivial variation over the surface of the embryo (see Fig. \ref{globalTiling}E) and it is informative to compare it with the coarse-grained inferred stress tensor Eq. \ref{stress}. In constructing the latter, it is helpful that GVM inference can be carried out locally on partially overlapping image frames and ``stitched" (see Appendix D for details) into a continuous coarse-grained stress field for the whole surface of the embryo (as illustrated in Fig. \ref{globalTiling}F). In Fig. \ref{globalTiling}D we compare the inferred coarse-grained stress tensor with the measured coarse-grained myosin tensor \cite{Streichan17}. Both myosin and inferred stress are enriched in the lateral ectoderm and are anisotropic along the DV axis, with the quantitative agreement at the level of about 60\% (which corresponds to average $\Delta \sim .4$ in Fig. \ref{globalTiling}F). In Fig. \ref{globalTiling}D we compare the inferred coarse-grained stress tensor with the measured coarse-grained myosin tensor \cite{Streichan17}. 

To more exactly quantify the difference between the measured myosin tensor $m^{ab}$ (as defined in \cite{Streichan17}, 
) and the inferred stress tensor $\sigma^{ab}$ we defined a normalized root-mean-square (r.m.s.) deviation: 
\begin{equation}\label{deltaEqn}
\Delta (r) \equiv \left [ \frac{  \langle  Tr \left( \mathbf{m} - \lambda \mathbf{\sigma} \right)^2 \rangle_r} {  \langle Tr (\mathbf{m}^2) \rangle_r }\right ]^{1/2}
\end{equation}
where $\mathbf{m},\mathbf{\sigma}$ abbreviate $m^{ab},\sigma^{ab}$ respectively and  $\langle ... \rangle_r$ stand in for averaging over a coarse-graining region at location $r$ and $\lambda$ is the unknown overall scaling factor relating myosin and stress which we chose so as to minimize the global average of $\Delta$. Hence $1-\Delta(r)$ is the measure of local agreement (in both magnitude and anisotropy) between of $\mathbf{\sigma}$ and $\mathbf{m}$ within a coarse-graining region $r$.

In general, the inferred stress tensor is consistent with the measured myosin tensor -- both exhibit strong anisotropy localized to the lateral ectoderm, with principal axis along the DV axis. The most substantial disagreement between the inferred stress and measured myosin tensors, as shown in Fig. \ref{globalTiling}F, is localized at the center of the dorsal side of the embryo; myosin is spatially inhomogeneous along the DV axis (high at the lateral sides) in contrast to the inferred stress tensor, observed to be constant along the DV axis. A plausible explanation for this discrepancy is that we are directly comparing the total myosin tensor, which is thought to drive early morphogenetic flow \cite{Streichan17} and thus is unbalanced, to our inferred stress tensor which explictly assumes mechanical equilibrium.

Our previous study \cite{Streichan17} related observed meso-scale myosin distribution $m^{ab}$ and observed morphogenetic flow, by focusing on the divergence of the myosin tensor $\nabla_a m^{ab}$ which corresponds to the ``unbalanced" internal stress within the tissue that generates cellular flow. (Note, following \cite{Streichan17} we assume $\sigma^{ab} \sim m^{ab}$.) However, only a fraction of myosin contributes the unbalanced stress, the rest generating internal stress obeying force balance: $m^{ab}=m_{U}^{ab}+m_B^{ab}$ with the ``balanced" fraction of myosin $m_B^{ab}$ being divergence-less, $\nabla^a m_B^{ab}=0$. It is the latter component of myosin that is expected to correlate with the predictions of GVM-based inference. 

To decompose the measured  myosin tensor \cite{Streichan17} into ``balanced" and ``unbalanced" components  we note that any 2D symmetric tensor can be  written as
\begin{equation}\label{tensorDecomp}
m^{ab} \equiv \left[\nabla^a u^b + \nabla^b u^a\right] + {\varepsilon^{ac}} {\varepsilon^{bd}} \nabla^c \nabla^d \varphi 
\end{equation}
(where $\varepsilon^{ac}=-\varepsilon^{ca}$ is the antisymmetric unit tensor). The divergence of the last term is zero and  it can be identified as $m_B^{ab}$, the balanced component of the myosin tensor. Taking the divergence of Eq. (\ref{tensorDecomp}) yields a partial differential equation for the vector field $u^b$:\begin{equation} \label{dynComponent}
 \nabla^2 u^b + \nabla^b \nabla^a u^a =\nabla^a m^{ab} 
\end{equation}
which was solved using the same method as in \cite{Streichan17}, see Methods. (Strictly speaking Eq. (\ref{dynComponent}) defined $u_a$ only up to a harmonic gradients $u_a \rightarrow u^a+\nabla^a \psi+\varepsilon^{ac}\nabla^c \omega$ with $\nabla^2 \psi=\nabla^2 \omega=0$. Since the only solution to the latter equations on a closed surface of genus zero is a constant, our solution for $u_a$ is unique.)  Eqns. (\ref{tensorDecomp}, \ref{dynComponent}) provide an explicit determination of the balanced component of myosin tensor:
\begin{equation}\label{balancedMyo}
m^{ab}_{B} \equiv m^{ab}-\left[\nabla^a u^b + \nabla^b u^a\right]
\end{equation}

\begin{figure}[ht] 
\centerline{
\includegraphics[width=.45\textwidth]{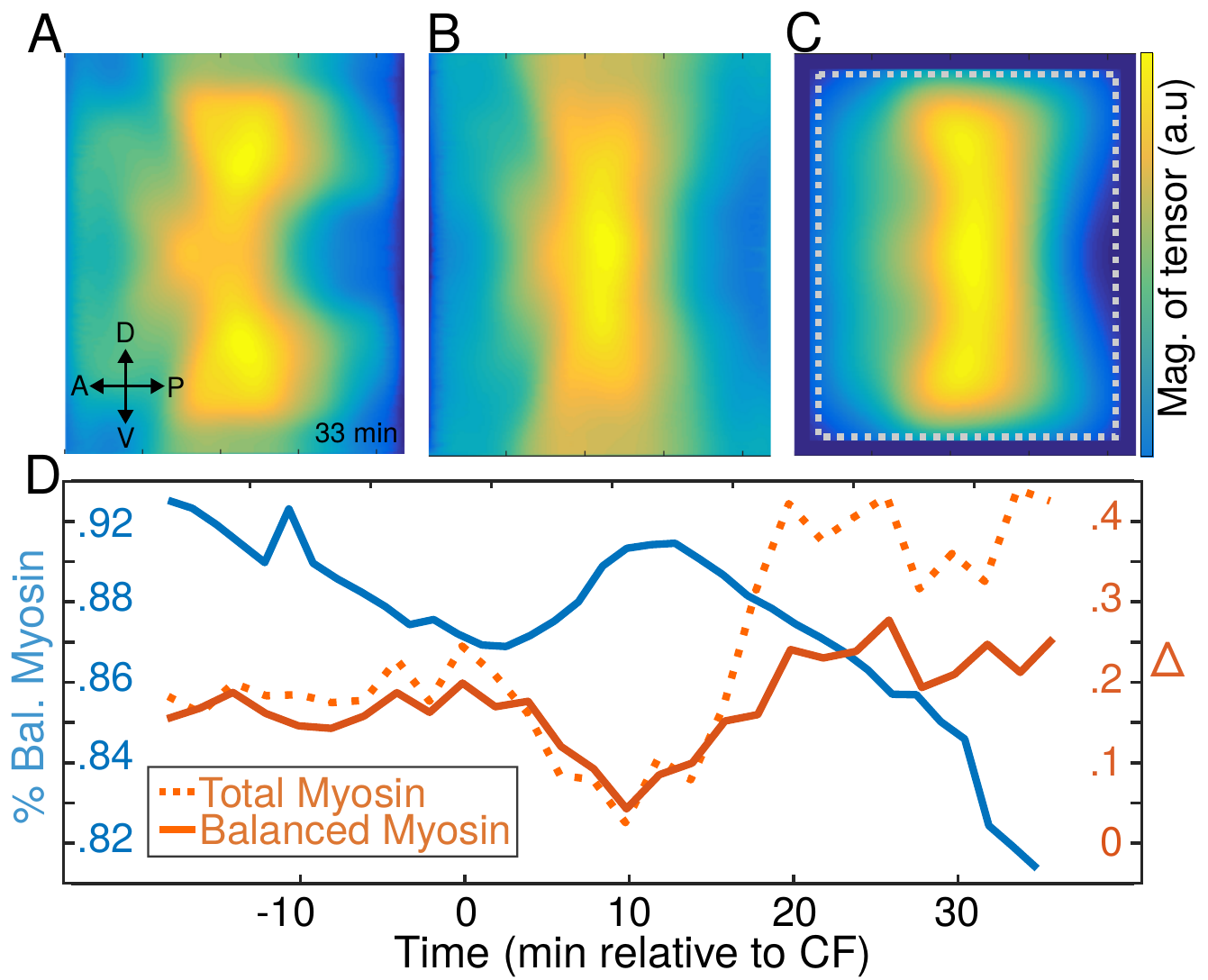}
}
\caption{\textbf{Balanced myosin versus total myosin.} A) The magnitude of the myosin tensor measured in \cite{Streichan17} 33 minutes after the formation of the ventral furrow. 
B) Same for the balanced component myosin tensor, defined by Eqn. (\ref{balancedMyo}), at the same  time. Note that the balanced myosin is approximately constant along the DV axis 
in contrast with the total myosin displayed in (A). 
C) The magnitude of the inferred stress tensor at the same time-point. Note greater similarity with (B) than with (A). The dotted gray box shows the region that was amenable to image segmentation. 
D) The average misalignment Eq. (\ref{deltaEqn}) between inferred stress and myosin tensors (balanced/total displayed as solid/dashed orange lines respectively). The balanced fraction of myosin is shown in blue: note that the surprisingly high fraction of myosin falling into the balanced component! As expected, the inferred stress tensor provides a better approximation for the balanced component of myosin tensor then for the total.  
}
\label{globalCompare}
\end{figure}

Fig. 5 displays the distribution of $m^{ab}_{B}$ on the surface of the embryo and compares it with the total myosin distribution and the inferred stress. As shown in Fig. \ref{globalCompare}D balanced myosin dominates accounting for more than $80\%$ of the total, but the unbalanced component increases with time, especially upon the onset of the GBE (10min post CF). We also find that removing the unbalanced component from the myosin distribution being compared to the inferred stress, substantially increases the agreement between the two (see Fig. \ref{globalCompare}D). 
Specifically, during GBE the alignment of our inferred stress tensor relative to the total myosin tensor decreases to $\sim\!60\%$, consistent with the higher fraction of unbalanced myosin and thus morphogenetic flow.

\begin{figure}[ht] 
\centerline{
\includegraphics[width=.5\textwidth]{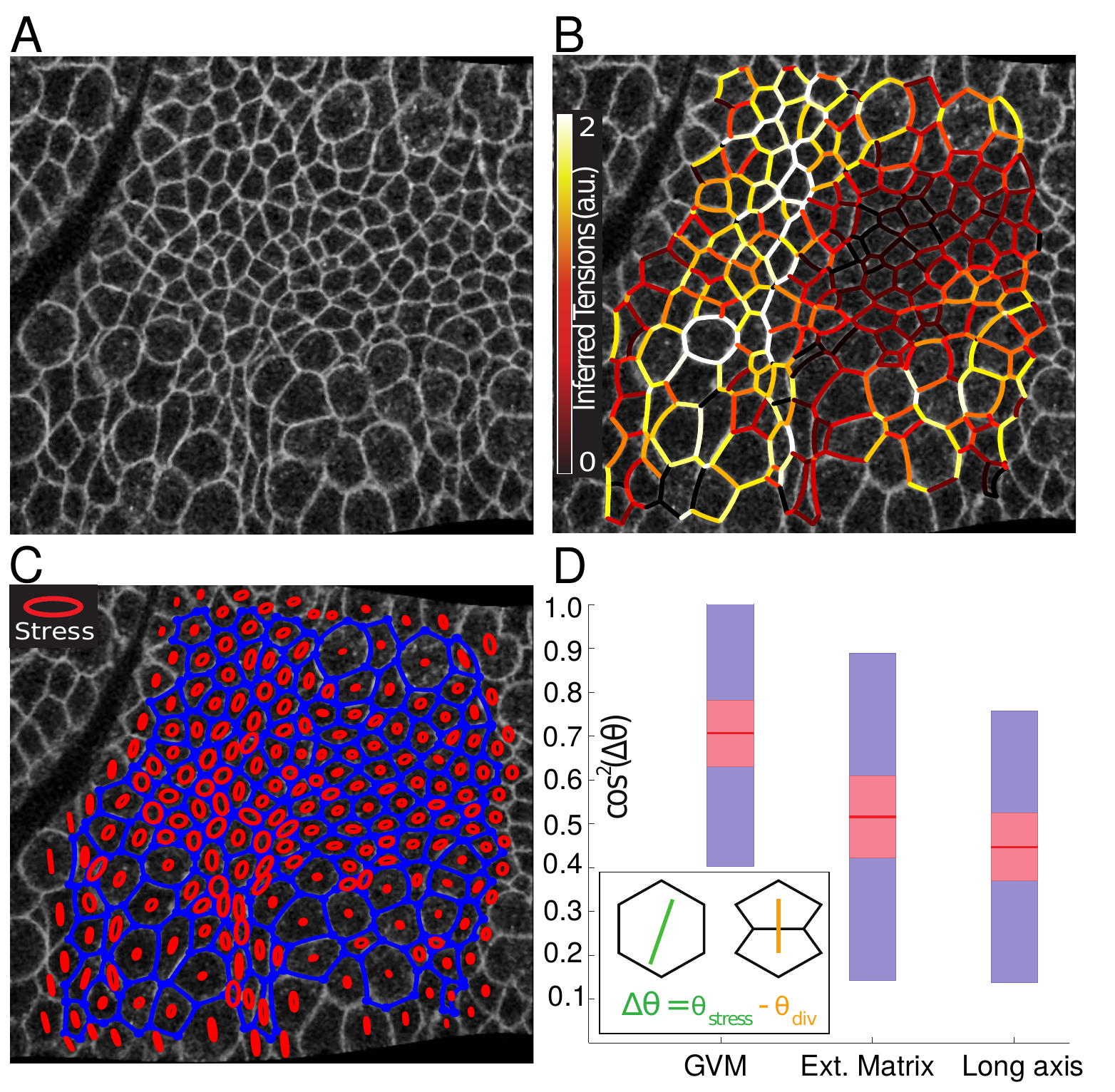}
}
\caption{ \textbf{Correlation of inferred stress and cell-division axis.} A) Confocal image of the \emph{Drosophila} lateral ectoderm, near the cephalic furrow, during the late phase of Germ Band Extension (GBE) at the onset of cell divisions, approximately 45 minutes after the formation of the cephalic furrow. 
B) An overlay of inferred tensions on the CAP array. 
C) Same as (B) but with the average stress tensor for each cell plotted as an ellipse. The major/minor axis of the ellipse corresponds to the principal axes of stress. 
D) Comparison of the cell-cleavage axis for cell division events (within mitotic domains 6 and 11  \cite{Foe89})
and the principal (extension) axis of inferred local stress  (using different methods). Stress  for mitotic cells was estimated 2 minutes before the registered time of division. While GVM-based analysis finds significant correlation, predictions based on the extended ``matrix inverse" \cite{Chiou12, CellFit} or cell elongation axis are not distinguishable from random.}
\label{localCompare}
\end{figure}

Another advantage to the GVM inference is it can be applied to cell arrays with arbitrarily large pressure differences. Let us now provide an example of how our variational stress inference can be applied to study interesting questions concerning mechanical control of biological phenomena in the context of cell division orientation. 

It is known that the spatio-temporal patterning and orientation of cell divisions plays an important role in morphogenesis. 
Mitotic domains of synchronously dividing cells partition the \emph{Drosophila} embryo in a highly regular manner that directly shape eventual larval segments \cite{Foe89}.
Additionally, the patterning of mitotic spindle orientation has been suggested to contribute to elongation of the posterior region of the lateral ectoderm \cite{Morais07}.
While the upstream signal that instructs orientation of cell cleavage plane is unknown, studies suggest that mechanical tension within the tissue contributes to spindle alignment \cite{Louveaux16,Nestor14} in the dividing cell. 

To test the hypothesis that cell cleavage axis tends to align with local tension within cells, we analyzed  70 tracked divisions in mitotic domains 6 and 11  (as defined in \cite{Foe89}) during the late phase of GBE (20-35min post CF).
Fig. \ref{localCompare}AB, provide an example of using the GVM method to infer tensions in individual interfaces of a cellular network.
The cell cleavage axis was compared directly to the orientation of the tension axis determined from the inferred stress tensor, Eq. (\ref{stress}). 
We found that cell cleavage indeed correlates strongly with inferred tension and that the principal axis of stress is a much better predictor of spindle orientation at the time of division than the  commonly used ``long axis" defined directly by cell elongation. Furthermore, the GVM-based inference  was more accurate  (based upon 95$\%$ confidence intervals using t-test) than the earlier ``matrix inverse" approach of \cite{Chiou12}. The improved accuracy is due to the GVM's ability to capture large pressure differentials between cells in a morphologically heterogeneous cell arrays, as exemplified by Fig. 5A.

\subsection*{Discussion}
The mechanical inference method described here was based on the model which assumed that (i) the 2D epithelial cell array is instantaneously in an approximate mechanical equilibrium, (ii) cell mechanics can be approximated by the balance of cytoskeletal tension localized at cell interfaces (and varying from one edge to another) and the effective areal pressure (preventing collapse of the apical surface of cells). Together, these assumptions place a non-trivial constraint on cell geometry that is readily testable on the basis of imaging data. The existence of local geometric constraints facilitated the formulation of a local mechanical inference scheme that combined estimation of stress with the simultaneous determination of the best-fit cell geometry from imaging data. The method was observed to be a significant improvement over similar methods, both at the scale of individual cells and the mesoscopic scale. 

Interestingly, equilibrium CAP tilings, which we used to approximate 2D epithelial cell geometry, are closely related to the familiar Voronoi tessellations of the plane. 
By virtue of its duality to a Delaunay triangulation, a Voronoi tiling satisfies the geometric compatibility constraint given by Eq. (\ref{genCompCond}). 
Equilibrium CAP tilings generalize the Voronoi construction to circular arcs instead of straight edges and correspondingly have a non-planar dual triangulation (see Appendix A. for the explicit construction). In the limit of constant pressure equilibrium CAP tilings reduce to standard Voronoi polygons with one additional degree of freedom per cell associated with the ``isogonal" deformation previously decribed in \cite{Noll17}.

In addition to testing the validity of the GVM-based stress inference, our analysis of myosin and inferred stress distributions  in {\it Drosophila} embryo has revealed that despite the dynamical nature of GBE, the epithelial shell of the embryo maintains approximate mechanical equilibrium, in the sense that mechanical stress associated with the observed myosin distribution is mostly (at the 80\% level) balanced internally and does not contribute to cellular flow. This conclusion is reached by a direct analysis of the measured myosin tensor, Eq. (\ref{balancedMyo}). Quite remarkably, the presence of this balanced internal stress is also correctly inferred from the GVM-based analysis of cell geometry across the surface of the embryo. The conclusion that tissue flow coexists with approximate internal force balance within a rearranging array of cells, provides an interesting insight into mechanics of tissues.

Also of biological interest in the analysis above, is the quantitative mapping of the unbalanced myosin, which according to \cite{Streichan17} acts as a driver of global morphogenetic flow. Disregarding the errors at the anterior and posterior poles that arise from image resolution issues, the largest deviation from mechanical balance is found (Fig. \ref{globalCompare} ABC) along the dorsal surface, where total myosin falls below the level needed to for internal force balance.
This suggests an important role for DV patterning in the convergent extension flow. We believe the ability to estimate global patterns of balanced and unbalanced stress on arbitrary two-dimensional surfaces opens up a novel method in which one can identify the factors that \emph{drive} morphological change.

We expect the GVM-based stress inference to be immediately useful for experimentalists studying tissue mechanics and the mechanics of morphogenesis of entire organs. 

\subsection*{Materials and Methods}

\paragraph*{Confocal microscopy}
Raw data shown in Fig. 4 was taken on a Leica SP8 confocal microscope equipped with two HyD detectors, a 40x / NA 1.1 water immersion objective, and 561 nm laser line. 
\paragraph*{Light sheet imaging}
In toto images for Figs 5 and 6 where taken on a custom-built multi view light sheet microscope described in \cite{Krzic2012}. Briefly, the setup consisted of two excitation and two detection arms. On each detection arm, the microscope was equipped with a water-dipping lens (Apo LWD 25x, NA 1.1, Nikon Instruments, Inc.), a filter wheel (HS-1032, Finger Lakes Instrumentation LLC) with emission filters (BLP02-561R-25, and BLP01-488R-25, Semrock, Inc), a tube lens (200 mm, Nikon Instruments, Inc.) and sCmos camera (Hamamatsu Flash 4.0 v3). In this way an effective pixel size of $.26\mu m$ was achieved. Illumination consisted of a water-dipping objective (CPI Plan Fluor 10x, NA .3, Nikon Instruments, Inc), a tube lens (same as above), a scan lens (S4LFT0061/065, Sill optics GmbH Co. KG), and a galvanometric scanner (6215h Cambridge Technology, Inc.). Illumination was based on laser lines (06-MLD 488nm, Cobolt AB, and 561 LS OBIS 561nm, Coherent, Inc.). 3D volumes where generates by translating samples through the sheet using a linear piezo stage (Physik Instrumente P-629.1CD controlled by C-867). Multiple views where generated by a rotation stage (Physik Instrumente U-628.03, C-867 controller), combined with a linear actuator (M-231.17, C-863 controller). Electronic control of the microscope was based on Micro Manager \cite{Edelstein2014}, and custom written MATLAB code. Fusion of individual views taken at 45 degrees angles was carried out using FIJI multi-view fusion plugins\cite{Preibisch2014}. Cartographic projections where generated using ImSAnE \cite{Heemskerk15}. 
\paragraph*{Fly stocks}
Sqh-GFP; membrane-mCherry 
\paragraph*{Numerical solution of Eq. (\ref{dynComponent})}
A 2D triangulated mesh of the embryo surface was constructed using ImSaNe \cite{Heemskerk15} and FELICITY \cite{Walker2018} -- a finite element software package for MATLAB -- was utilized to solve the PDE and compute surface derivatives. We refer the reader to \cite{Streichan17} for a detailed description of the method.
\paragraph*{Image segmentation}
In-vivo data was segmented using a custom pipeline implemented in MATLAB, available on Github at {\it nnoll/TissueAnalysisSuite}. 
{\it ilastik}, a supervised machine learning classifier, was used to as a pre-processing step for each image, followed by the application of a Laplacian of Gaussian filter. 
The resultant image was segmented using the watershed algorithm.

Vertices were defined as branch points of the resultant skeletonization -- edges are segmented as the set of boundary pixels that run between two such branch points. 
Our CAP is parameterized by not only vertex position but also edge curvature. 
Each edge was fit to a circular arc using the Pratt method, which is robust for small angle samplings of the underlying circle. 
Interfacial myosin concentration was measured by dilation of each segmented edge by 2 pixels and averaging over the resulting set of pixels in the myosin channel.
All segmentation information is stored within a custom data structure and can be immediately used for the GVM inference.

Cell divisions during late germ band extension were registered by tracking cells. 
Cell tracking was achieved by computing pixel overlaps between segmented cells in subsequent time points -- cells were paired based upon the cell they most overlap with in the succeeding frame. 
Mitotic cells were defined as tracking events where two cells overlapped with one in the previous time point.
The tracking was manually curated to ensure no false divisions were called. 

\paragraph*{3D reconstruction}
ImSaNe \cite{Heemskerk15} was used to measure, parameterize, and store the surface and embedding coordinates of the Drosophila embryonic surface. 
Segmentation of cells was done using the cylindrical mapping of the embyro. 
The 3D vertex positions were subsequently estimated using the embedding grids obtained from the ImSaNe algorithm. 

We computed the mesoscopic myosin distribution in the same way as detailed in \cite{Streichan17}. 
The output from the automated segmentation of myosin is a summation over microscopic nematic tensors of the form
\begin{equation}
    m^{ab}(\bm{r}) = \displaystyle\sum\limits_{\langle i,j\rangle>} m_{ij} \hat{n}_{ij}^a \hat{n}^b_{ij} \delta^2(\bm{r}-\bm{r}_{ij})
\end{equation}
This was averaged using a Gaussian filter.

\begin{acknowledgments}
The authors gratefully acknowledge stimulating discussions with Madhav Mani, Idse Heemskerk, and Eric Wieschaus. This work was supported by the NSF PHY-1220616 (BIS) and NICHD 5K99HD088708-02 (SJS).
\end{acknowledgments}

\appendix

\renewcommand{\thefigure}{A.\arabic{figure}}
\setcounter{figure}{0}

\section{The generalized Voronoi construction and the geometric interpretation of unconstrained variables }
In this section we show that variables $\{\bm{q}_\alpha\}$, introduced in Eq. (\ref{centroidDecomp}),  can be thought of as vertices `dual' to cells of the CAP tessellation that, along with the introduced weights $\{z_\alpha,p_\alpha\}$, generalize the passage between Voronoi tessellations and Delaunay triangulations. 
In analogy with the Voronoi construction,  given a pair of triangulation vertices $\bm{q}_\alpha, \bm{q}_\beta$, we define an edge $\alpha\beta$ of the corresponding `dual' cell as the locus of points equidistant from $\bm{q}_\alpha$ and $\bm{q}_\beta$, so that $d_\alpha^2 (r)= d_\beta^2 (r)$ with a generalized "distance" defined by
\begin{equation}\label{metricDef}
d_\alpha^2(\bm{r}) = p_\alpha|\bm{r} - \bm{q}_\alpha|^2 + p_\alpha z^2_\alpha 
\end{equation}
Thus, points on the edge $\bm{r}_{\alpha \beta}$ satisfy 
\begin{equation}\label{edgeDef}
p_{\alpha} \, |\bm{r}_{\alpha\beta}-\bm{q}_{\alpha}|^2 + p_\alpha z^2_{\alpha}  = p_{\beta} \, |\bm{r}_{\alpha\beta}-\bm{q}_{\beta}|^2 + p_\beta z^2_{\beta}
\end{equation}
which defines a circular arc with centroid  $\bm{\rho}_{\alpha\beta} \equiv \frac{p_{\alpha}\bm{q}_{\alpha} - p_{\beta}\bm{q}_{\beta}}{p_{\alpha} - p_{\beta}}$ the  radius $R_{\alpha\beta}$ defined by Eq. (\ref{tensionDef}). Interpreting $p_{\alpha}$ as the `apical' $2$D pressure immediately implies that the tension of this edge is 
\begin{equation} \label{tensionEq}
T_{\alpha\beta} = \sqrt{p_{\alpha}p_{\beta} |\bm{q}_\alpha - \bm{q}_\beta|^2 - \left(p_{\alpha}-p_{\beta}\right)\left[p_\alpha z^2_\alpha - p_\beta z^2_\beta\right]}
\end{equation}

We note that in the limit that pressure $p_\alpha = p_0$ is set to a uniform constant, $R_{\alpha\beta}$ diverge, turning arcs into straight lines and we recover the properties of the ATN model presented in \cite{Noll17}, where $z^2_\alpha$ plays the role of the ``isogonal" soft mode of cell deformation. Additionally, setting all $z_\alpha = z_0$ recovers the usual Delaunay to Voronoi duality.

\section{Initial condition for the GVM}

The initial condition for the variational parameters used in the minimization of Eq. (\ref{inverseEnergy}) is obtained piecewise; we first approximate $\{\bm{q}_\alpha,p_\alpha\}$ independent of $z_\alpha$ by fitting edge centroids $\bm{\rho}_{\alpha\beta}$.
As depicted in Fig. \ref{forwardDef}, the vector pointing from each arc's centroid to either of the attached vertices, $\bm{r}_{i}-\bm{\rho}_{\alpha\beta}$, is orthogonal to the local tangent direction along the ${\bm r}_{\alpha \beta}$ edge.
We denote the measured tangent of edge $\alpha\beta$ at vertex $i$ as $\hat{\bm{\tau}}_{i,\alpha\beta}$. 

The above geometric constraint can be imposed variationally on our parameters.
It is convenient for the following discussion to introduce the shorthand notation $\bm{d}_{i,\alpha\beta} \equiv  \bm{r}_i -\frac{1}{p_\alpha - p_\beta} \left[ p_\alpha \bm{q}_\alpha - p_\beta\bm{q}_\beta\right]$
Initial estimates for $q_{\alpha},p_\alpha$ are thus obtained by minimization of
\begin{equation}
\mathcal{E} = \frac{1}{n_e} \!\displaystyle\sum\limits_{<\alpha,\beta>}\!\! \left[\bm{d}_{i,\alpha\beta}\!\cdot\! \bm{\hat{\tau}}_{i,\alpha\beta}\right]^2\! + \! \left[\bm{d}_{j,\alpha\beta} \!\cdot\! \bm{\hat{\tau}}_{j,\alpha\beta}\right]^2 
\end{equation}
The solution is constrained such that the average magnitude of $|\bm{d}_{i,\alpha\beta}|$ equals the averaged measured radius of curvature in the image to ensure we don't infer the trivial solution ($\bm{q}_\alpha = \bm{q}_\beta, p_\alpha = p_\beta$).

Initial estimates of $z_\alpha^2$ is obtained from the variational parameters obtained above by inverting the linear set of equations defined by the square of Eq. (\ref{tensionEq}). 
\begin{equation}
\frac{p_{\alpha} p_{\beta}|\bm{q}_\alpha-\bm{q}_\beta|^2 - R_{\alpha\beta}^2}{p_\alpha-p_\beta} = p_\beta z^2_\beta - p_\alpha z^2_\alpha
\end{equation}
subject to the constraint that $\sum_{\alpha} p_\alpha z^2_\alpha = 1$. $R_{\alpha\beta}$ is segmented from the image.

\section{In-silico benchmark of GVM algorithm}
\begin{figure}[ht]
\centerline{
\includegraphics[width=.5\textwidth]{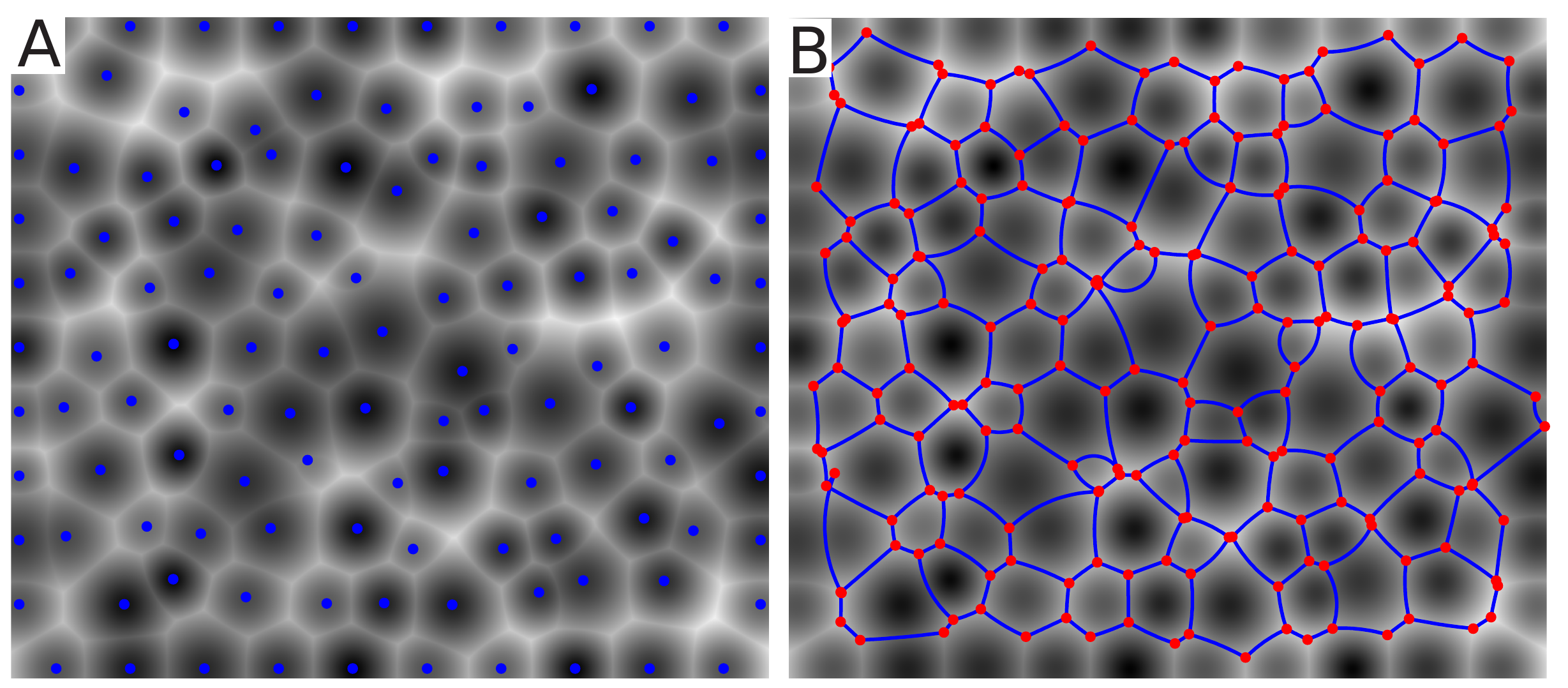}
}
\caption{A) An example plot of the image of weighted distance from each generating point produced using MATLAB's bwdist function and the algorithm described. 
The location of the generating points $\bm{q}_{\alpha}$ are shown as blue dots.  
B) The resultant CAP network.
The segmentation output from the watershed algorithm is used to infer the triangulation topology. 
This is then passed into our forward equations to generate the \emph{exact} positions of vertices and curvatures of edges. The result, using the original values for the generating points and the measured triangulation topology, is overlayed in blue over the original distance map.}
\label{simEg}
\end{figure}
To test the robustness of the GVM inference, synthetic data was generated by initializing a triangular lattice of $\sim 120$ generating points $\bm{q}_{\alpha}$ within a rectangle of size $[1,\sqrt{3}/2]$.
Using the Delaunay/Voronoi correspondence outlined in Appendix A, one can easily generate arbitrary CAP networks -- i.e. for any set $\{\bm{q}_\alpha,p_\alpha,z_\alpha\}$ -- in mechanical equilibrium. 
Edges within the CAP network are defined by Eq. (\ref{edgeDef}) and thus $d^2_\alpha$ must be computed from $\{\bm{q}_\alpha\}$.
Distance from $\bm{q}_{\alpha}$ is calculated using MATLAB's bwdist function and then scaled according to Eq. (\ref{metricDef}) to obtain $d_{\alpha}^2(\bm{r})$. 
This procedure is repeated for each point $\bm{q}$.
The minimum of $\{d^2_\alpha\}$ for each spatial location is taken - the net result is a scalar field that measures the minimum weighted distance away from any triangulation vertex, $d^2(\bm{r}) = \text{min}_{\alpha}\, d_{\alpha}^2(\bm{r})$. 
An example of the above procedure is shown in Fig. \ref{simEg}A. 
Edges are `ridges' representing local maxima $d^2(\bm{r})$ that are found easily using the watershed algorithm -- Fig. \ref{simEg}B shows an example.
The resultant equilibrium network can then be immediately calculated from the original parameters for $\{\bm{q}_{\alpha}\}$, $\{p_{\alpha}\}$, and $\{z_{\alpha}\}$. 

\begin{figure}[ht]
\centerline{
\includegraphics[width=.5\textwidth]{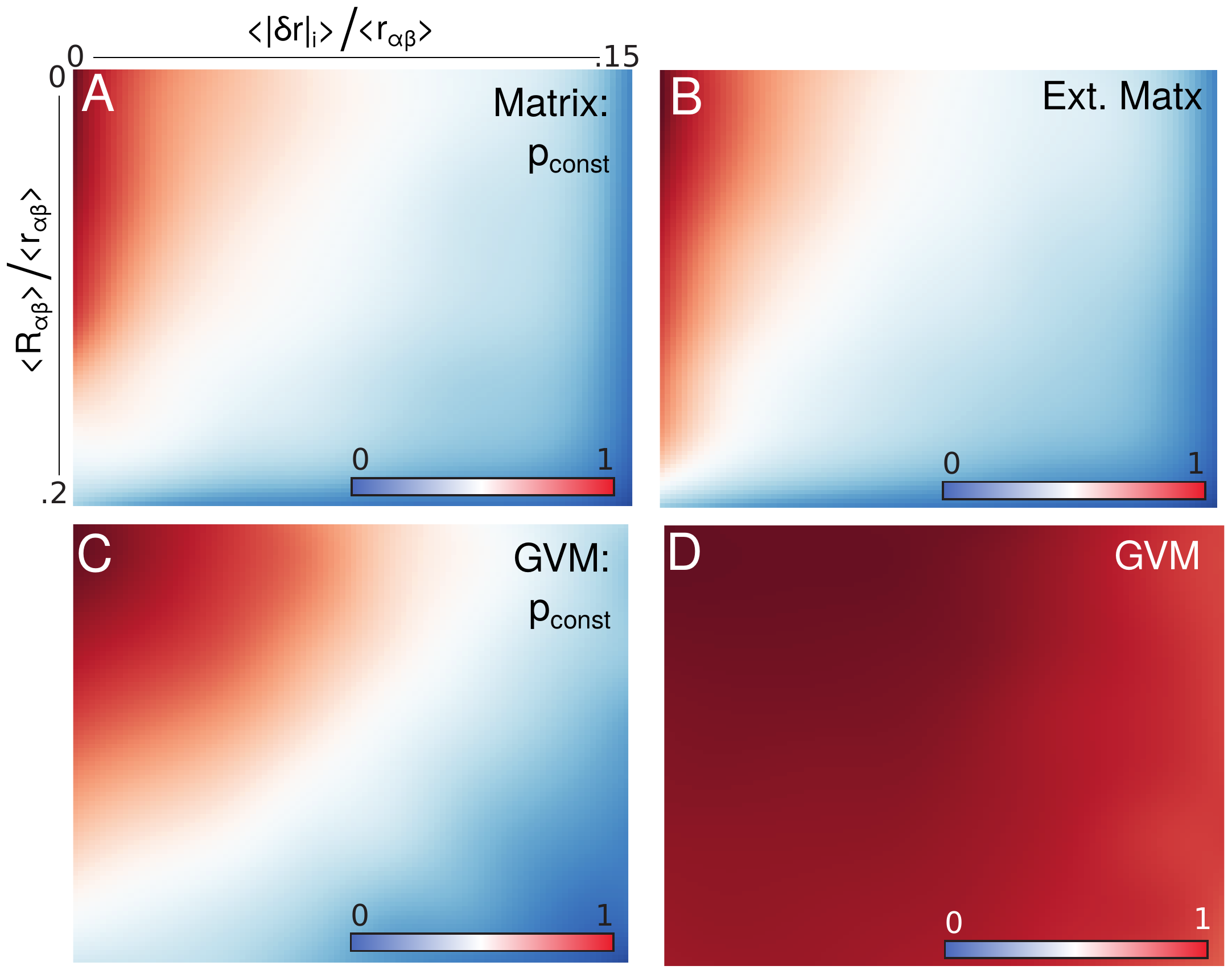}
}
\caption{Correlation between inferred tensions and pressures against the known syntetic values -- displayed as a function of average edge curvature $R_{\alpha\beta}$ (due to pressure inhomogeneity) in units of chord length $r_{\alpha\beta}$, on the vertical axis, and average noise added to vertex and edge centroid positions $|\delta \bm{r}_i|$ (also in units of average chord length $|\bm{r}_{\alpha\beta}$ and) shown on the horizontal axis.  Results were tested for four variants of mechanical inverse algorithms: (A) matrix pseudo-inverse assuming constant tension, (B) the extended matrix inverse, augmented with rows enforcing the measured curvature, (C) a constrained GVM based method that infers mechanics subject to constant tension and (D) the GVM inference. All estimates shown were averaged over $100$ randomly generated cell array for each parameter value.}
\label{inverseSensitivity}
\end{figure}

Intercellular pressures $p_{\alpha}$ were sampled from a uniform distribution to generate lattices of varying curvature. 
Additionally, values for $\{z_{\alpha}\}$ were pulled from a Gaussian with standard deviation $.2$. 
White noise, denoted $\bm{\delta}$ in Fig.\ref{inverseSensitivity}, was added to both vertex position $\bm{r}_{i}$ as well as position of edge centroids $\bm{\rho}_{\alpha\beta}$ in order to assay sensitivity of the algorithm. 

We benchmarked the relative efficacies of four mechanical inverse algorithms as a function of measurement noise and the contribution of pressure to the mechanical balance: (A) the matrix inverse with constant pressure \cite{Chiou12}, (B) the extended matrix inverse of \cite{CellFit}, (C) a variant of the GVM inference constrained to constant cell pressures, (D) the GVM algorithm. 
The resultant correlation between inferred and the known generated tensions and pressures are shown in Fig. \ref{inverseSensitivity}(A-D). 
Both the variational constant pressure (C) and GVM inferences (D) are significantly less sensitive to measurement noise than their matrix inverse counterparts (algorithms (A) and (B) respectively). 
Furthermore, we see algorithm (B) loses accuracy at moderate curvature values (shown on the vertical axis, normalized to the average chord length between vertices) due to the assumptions of small pressure differentials required to linearize the underlying force-balance equations. 
Conversely, GVM is robust to both, showing high correlation in the entire tested parameter regime.

\section{GVM algorithm applied to curved surfaces}
As proposed, the GVM can be easily extended to formulate a tractable inference scheme for the balanced mechanical stress within a curved tissue's tangent plane. 
Due to the inclusion of edge curvature information, the inverse is extensively over-determined which allows one to simultaneously infer both bulk and boundary stress using information of just bulk geometry. 
Consequently, the global mechanical state can be `stitched' together by inferring stress on local patches of cells that can, with good approximation, be treated as planar.
\begin{figure}[ht]
\centerline{
\includegraphics[width=.5\textwidth]{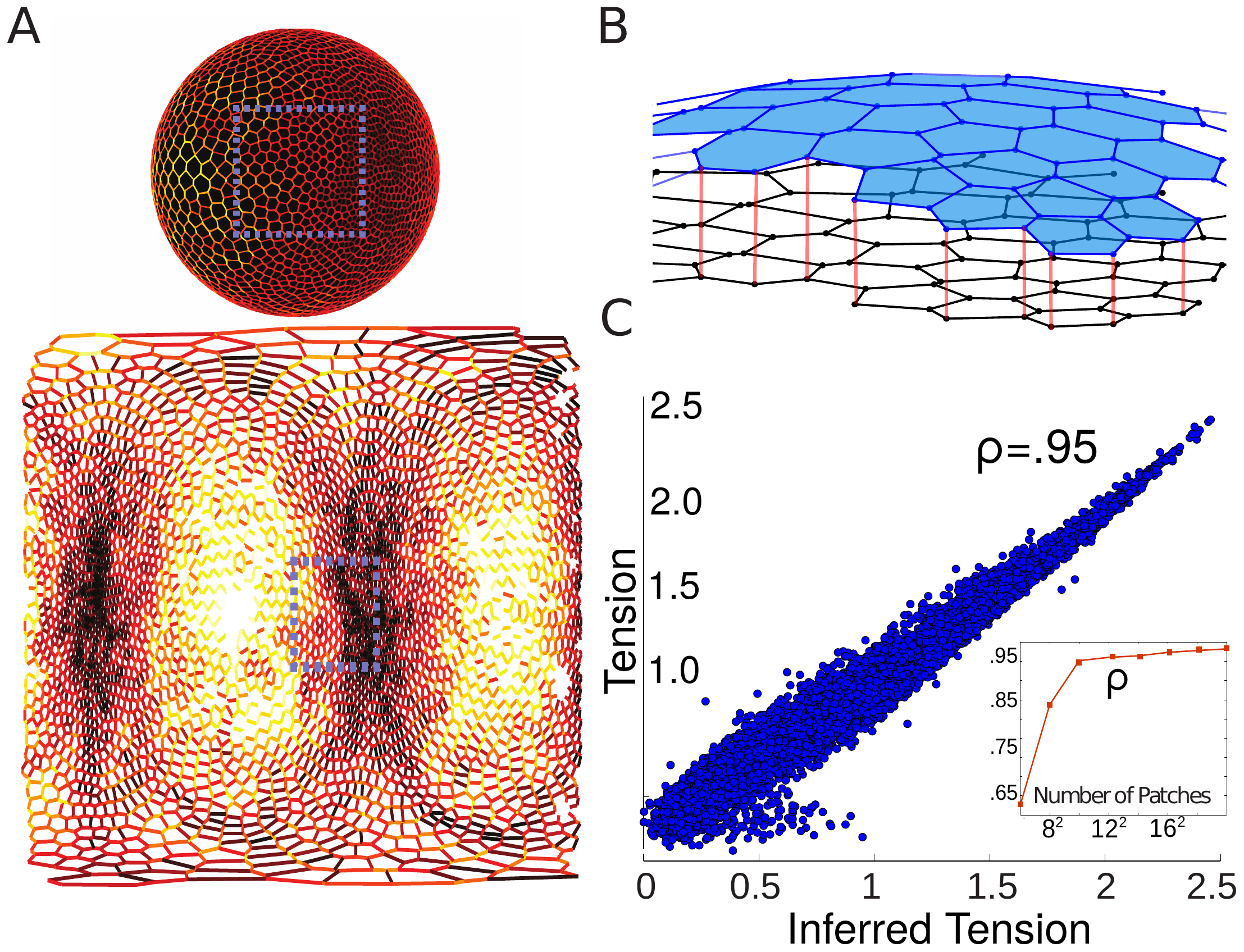}
}
\caption{
(A) A synthetic spherical embryo of $\sim 3000$ cells, plotted both in the embedded space, as well as the cylindrical unwrapping analogous to the embryonic data shown in the main text. 
Simulated interfacial tensions are plotted as a heatmap in both domains. 
(B) Graphical depiction of the process used to infer stresses in the tangent plane of a small region of cells. 
All cells are segmented in 3D using ImSaNe \cite{Heemskerk15}
Each local cellular patch is projected from the 3D embedding space onto the 2D plane that minimizes the sum of squares of deviation.
The resulting projection is used as an input into the GVM method.
(C) Scatter plot between the inferred tensions using the workflow outlined and the known tensions shown in (A). 
The inset displays the dependence on the number of patches used to cover the sphere. 
As was expected, correlation is monotonic with sampling resolution.  
}
\label{curvedInverse}
\end{figure}
The blue cell array depicted in Fig. \ref{curvedInverse}B denotes the `true' apical surface of the epithelial tissue. 
Provided the area of interest is much smaller than the surface's radius of curvature, we can fit a well defined tangent plane to the patch. 
Let $\bm{R}_i^n$ denote the 3D position of the $i^{th}$ vertex within patch $n$ -- it is a matrix of size $3$ x $v_n$ where $v_n$ is the number of vertices contained in patch $n$. 
The best fit triad of vectors is obtained easily via an SVD decomposition $ \bm{R}_i^n = U_n \Sigma V^{T}_n$

The approximate planar graph of patch $n$, shown as the black cell array in Fig. \ref{curvedInverse}B, is obtained by projecting $\bm{R}_i^n$ onto the two principal components; GVM is applied on the distortion.
Importantly, the set of inferred tensions and pressures within each patch is unique up to an overall scale and thus there exists an undetermined relative scale between each patch - denoted $\lambda_n$.
We fix such scales by definining each patch to overlap by $1/4$ their linear extent such that a subset of edges are involved in multiple inferred regions. 
Hence, the scale $\lambda_{n}$ for each patch is found by minimizing the squared difference between inferred tensions of edges shared by adjacent patches globally, subject to the constraint that the average scale is $1$ to ensure a non-trivial solution $\lambda_{n} \ne 0$. 

This was used to define the patch size used in the empirical measurements during Drosophila embryogenesis. 
The outlined procedure was validated in-silico for synthetic spherical embryos containing roughly $3000$ cells, with mechanics patterned by a vertex model minimized on the surface of a sphere. 
An example of a simulated embryo with azimuthal pattern of tension is shown in Fig.\ref{curvedInverse}A, both in the embedding space and in the cylindrical unwrapping of the sphere, analogous to the data shown for the Drosophila embryo. 
As shown in Fig. \ref{curvedInverse}C, excellent agreement between the inferred and known tensions was found \emph{provided} the patch size was small compared to surface curvature. 
The inset shows this occurred when the defined patches contained $100$ or less cells. 
This was used to define the patch size used in the empirical measurements during Drosophila embryogenesis.

\bibliography{refs}{}

\begin{thebibliography}{42}%
\makeatletter
\providecommand \@ifxundefined [1]{%
 \@ifx{#1\undefined}
}%
\providecommand \@ifnum [1]{%
 \ifnum #1\expandafter \@firstoftwo
 \else \expandafter \@secondoftwo
 \fi
}%
\providecommand \@ifx [1]{%
 \ifx #1\expandafter \@firstoftwo
 \else \expandafter \@secondoftwo
 \fi
}%
\providecommand \natexlab [1]{#1}%
\providecommand \enquote  [1]{``#1''}%
\providecommand \bibnamefont  [1]{#1}%
\providecommand \bibfnamefont [1]{#1}%
\providecommand \citenamefont [1]{#1}%
\providecommand \href@noop [0]{\@secondoftwo}%
\providecommand \href [0]{\begingroup \@sanitize@url \@href}%
\providecommand \@href[1]{\@@startlink{#1}\@@href}%
\providecommand \@@href[1]{\endgroup#1\@@endlink}%
\providecommand \@sanitize@url [0]{\catcode `\\12\catcode `\$12\catcode
  `\&12\catcode `\#12\catcode `\^12\catcode `\_12\catcode `\%12\relax}%
\providecommand \@@startlink[1]{}%
\providecommand \@@endlink[0]{}%
\providecommand \url  [0]{\begingroup\@sanitize@url \@url }%
\providecommand \@url [1]{\endgroup\@href {#1}{\urlprefix }}%
\providecommand \urlprefix  [0]{URL }%
\providecommand \Eprint [0]{\href }%
\providecommand \doibase [0]{http://dx.doi.org/}%
\providecommand \selectlanguage [0]{\@gobble}%
\providecommand \bibinfo  [0]{\@secondoftwo}%
\providecommand \bibfield  [0]{\@secondoftwo}%
\providecommand \translation [1]{[#1]}%
\providecommand \BibitemOpen [0]{}%
\providecommand \bibitemStop [0]{}%
\providecommand \bibitemNoStop [0]{.\EOS\space}%
\providecommand \EOS [0]{\spacefactor3000\relax}%
\providecommand \BibitemShut  [1]{\csname bibitem#1\endcsname}%
\let\auto@bib@innerbib\@empty
\bibitem [{\citenamefont {Farhadifar}\ \emph {et~al.}(2007)\citenamefont
  {Farhadifar}, \citenamefont {R{\"o}per}, \citenamefont {Aigouy},
  \citenamefont {Eaton},\ and\ \citenamefont {J{\"u}licher}}]{Far07}%
  \BibitemOpen
  \bibfield  {author} {\bibinfo {author} {\bibfnamefont {Reza}\ \bibnamefont
  {Farhadifar}}, \bibinfo {author} {\bibfnamefont {Jens-Christian}\
  \bibnamefont {R{\"o}per}}, \bibinfo {author} {\bibfnamefont {Benoit}\
  \bibnamefont {Aigouy}}, \bibinfo {author} {\bibfnamefont {Suzanne}\
  \bibnamefont {Eaton}}, \ and\ \bibinfo {author} {\bibfnamefont {Frank}\
  \bibnamefont {J{\"u}licher}},\ }\bibfield  {title} {\enquote {\bibinfo
  {title} {The influence of cell mechanics, cell-cell interactions, and
  proliferation on epithelial packing},}\ }\href@noop {} {\bibfield  {journal}
  {\bibinfo  {journal} {Current Biology}\ }\textbf {\bibinfo {volume} {17}},\
  \bibinfo {pages} {2095--2104} (\bibinfo {year} {2007})}\BibitemShut {NoStop}%
\bibitem [{\citenamefont {Rauzi}\ \emph {et~al.}(2008)\citenamefont {Rauzi},
  \citenamefont {Verant}, \citenamefont {Lecuit},\ and\ \citenamefont
  {Lenne}}]{Rauz08}%
  \BibitemOpen
  \bibfield  {author} {\bibinfo {author} {\bibfnamefont {Matteo}\ \bibnamefont
  {Rauzi}}, \bibinfo {author} {\bibfnamefont {Pascale}\ \bibnamefont {Verant}},
  \bibinfo {author} {\bibfnamefont {Thomas}\ \bibnamefont {Lecuit}}, \ and\
  \bibinfo {author} {\bibfnamefont {Pierre-Fran{\c{c}}ois}\ \bibnamefont
  {Lenne}},\ }\bibfield  {title} {\enquote {\bibinfo {title} {Nature and
  anisotropy of cortical forces orienting drosophila tissue morphogenesis},}\
  }\href@noop {} {\bibfield  {journal} {\bibinfo  {journal} {Nature cell
  biology}\ }\textbf {\bibinfo {volume} {10}},\ \bibinfo {pages} {1401}
  (\bibinfo {year} {2008})}\BibitemShut {NoStop}%
\bibitem [{\citenamefont {Blanchard}\ \emph {et~al.}(2009)\citenamefont
  {Blanchard}, \citenamefont {Kabla}, \citenamefont {Schultz}, \citenamefont
  {Butler}, \citenamefont {Sanson}, \citenamefont {Gorfinkiel}, \citenamefont
  {Mahadevan},\ and\ \citenamefont {Adams}}]{Blanchard09}%
  \BibitemOpen
  \bibfield  {author} {\bibinfo {author} {\bibfnamefont {Guy~B}\ \bibnamefont
  {Blanchard}}, \bibinfo {author} {\bibfnamefont {Alexandre~J}\ \bibnamefont
  {Kabla}}, \bibinfo {author} {\bibfnamefont {Nora~L}\ \bibnamefont {Schultz}},
  \bibinfo {author} {\bibfnamefont {Lucy~C}\ \bibnamefont {Butler}}, \bibinfo
  {author} {\bibfnamefont {Benedicte}\ \bibnamefont {Sanson}}, \bibinfo
  {author} {\bibfnamefont {Nicole}\ \bibnamefont {Gorfinkiel}}, \bibinfo
  {author} {\bibfnamefont {L}~\bibnamefont {Mahadevan}}, \ and\ \bibinfo
  {author} {\bibfnamefont {Richard~J}\ \bibnamefont {Adams}},\ }\bibfield
  {title} {\enquote {\bibinfo {title} {Tissue tectonics: morphogenetic strain
  rates, cell shape change and intercalation},}\ }\href@noop {} {\bibfield
  {journal} {\bibinfo  {journal} {Nature methods}\ }\textbf {\bibinfo {volume}
  {6}},\ \bibinfo {pages} {458} (\bibinfo {year} {2009})}\BibitemShut {NoStop}%
\bibitem [{\citenamefont {Etournay}\ \emph {et~al.}(2016)\citenamefont
  {Etournay}, \citenamefont {Merkel}, \citenamefont {Popovi{\'c}},
  \citenamefont {Brandl}, \citenamefont {Dye}, \citenamefont {Aigouy},
  \citenamefont {Salbreux}, \citenamefont {Eaton},\ and\ \citenamefont
  {J{\"u}licher}}]{Etournay16}%
  \BibitemOpen
  \bibfield  {author} {\bibinfo {author} {\bibfnamefont {Raphael}\ \bibnamefont
  {Etournay}}, \bibinfo {author} {\bibfnamefont {Matthias}\ \bibnamefont
  {Merkel}}, \bibinfo {author} {\bibfnamefont {Marko}\ \bibnamefont
  {Popovi{\'c}}}, \bibinfo {author} {\bibfnamefont {Holger}\ \bibnamefont
  {Brandl}}, \bibinfo {author} {\bibfnamefont {Natalie~A}\ \bibnamefont {Dye}},
  \bibinfo {author} {\bibfnamefont {Beno{\^\i}t}\ \bibnamefont {Aigouy}},
  \bibinfo {author} {\bibfnamefont {Guillaume}\ \bibnamefont {Salbreux}},
  \bibinfo {author} {\bibfnamefont {Suzanne}\ \bibnamefont {Eaton}}, \ and\
  \bibinfo {author} {\bibfnamefont {Frank}\ \bibnamefont {J{\"u}licher}},\
  }\bibfield  {title} {\enquote {\bibinfo {title} {Tissueminer: a multiscale
  analysis toolkit to quantify how cellular processes create tissue
  dynamics},}\ }\href@noop {} {\bibfield  {journal} {\bibinfo  {journal}
  {Elife}\ }\textbf {\bibinfo {volume} {5}},\ \bibinfo {pages} {e14334}
  (\bibinfo {year} {2016})}\BibitemShut {NoStop}%
\bibitem [{\citenamefont {Ingber}(2005)}]{Ingber05}%
  \BibitemOpen
  \bibfield  {author} {\bibinfo {author} {\bibfnamefont {Donald~E}\
  \bibnamefont {Ingber}},\ }\bibfield  {title} {\enquote {\bibinfo {title}
  {Mechanical control of tissue growth: function follows form},}\ }\href@noop
  {} {\bibfield  {journal} {\bibinfo  {journal} {Proceedings of the National
  Academy of Sciences}\ }\textbf {\bibinfo {volume} {102}},\ \bibinfo {pages}
  {11571--11572} (\bibinfo {year} {2005})}\BibitemShut {NoStop}%
\bibitem [{\citenamefont {Shraiman}(2005)}]{Shraiman05}%
  \BibitemOpen
  \bibfield  {author} {\bibinfo {author} {\bibfnamefont {Boris~I.}\
  \bibnamefont {Shraiman}},\ }\bibfield  {title} {{\selectlanguage
  {english}\enquote {\bibinfo {title} {Mechanical feedback as a possible
  regulator of tissue growth},}\ }}\href {\doibase 10.1073/pnas.0404782102}
  {\bibfield  {journal} {\bibinfo  {journal} {Proceedings of the National
  Academy of Sciences}\ }\textbf {\bibinfo {volume} {102}},\ \bibinfo {pages}
  {3318--3323} (\bibinfo {year} {2005})}\BibitemShut {NoStop}%
\bibitem [{\citenamefont {Pan}\ \emph {et~al.}(2016)\citenamefont {Pan},
  \citenamefont {Heemskerk}, \citenamefont {Ibar}, \citenamefont {Shraiman},\
  and\ \citenamefont {Irvine}}]{Pan16}%
  \BibitemOpen
  \bibfield  {author} {\bibinfo {author} {\bibfnamefont {Yuanwang}\
  \bibnamefont {Pan}}, \bibinfo {author} {\bibfnamefont {Idse}\ \bibnamefont
  {Heemskerk}}, \bibinfo {author} {\bibfnamefont {Consuelo}\ \bibnamefont
  {Ibar}}, \bibinfo {author} {\bibfnamefont {Boris~I}\ \bibnamefont
  {Shraiman}}, \ and\ \bibinfo {author} {\bibfnamefont {Kenneth~D}\
  \bibnamefont {Irvine}},\ }\bibfield  {title} {\enquote {\bibinfo {title}
  {Differential growth triggers mechanical feedback that elevates hippo
  signaling},}\ }\href@noop {} {\bibfield  {journal} {\bibinfo  {journal}
  {Proceedings of the National Academy of Sciences}\ }\textbf {\bibinfo
  {volume} {113}},\ \bibinfo {pages} {E6974--E6983} (\bibinfo {year}
  {2016})}\BibitemShut {NoStop}%
\bibitem [{\citenamefont {Frantz}\ \emph {et~al.}(2010)\citenamefont {Frantz},
  \citenamefont {Stewart},\ and\ \citenamefont {Weaver}}]{Frantz2010}%
  \BibitemOpen
  \bibfield  {author} {\bibinfo {author} {\bibfnamefont {Christian}\
  \bibnamefont {Frantz}}, \bibinfo {author} {\bibfnamefont {Kathleen~M.}\
  \bibnamefont {Stewart}}, \ and\ \bibinfo {author} {\bibfnamefont
  {Valerie~M.}\ \bibnamefont {Weaver}},\ }\bibfield  {title} {{\selectlanguage
  {english}\enquote {\bibinfo {title} {The extracellular matrix at a glance},}\
  }}\href {\doibase 10.1242/jcs.023820} {\bibfield  {journal} {\bibinfo
  {journal} {J Cell Sci}\ }\textbf {\bibinfo {volume} {123}},\ \bibinfo {pages}
  {4195--4200} (\bibinfo {year} {2010})}\BibitemShut {NoStop}%
\bibitem [{\citenamefont {Hartsock}\ and\ \citenamefont
  {Nelson}(2008)}]{AJ_review}%
  \BibitemOpen
  \bibfield  {author} {\bibinfo {author} {\bibfnamefont {Andrea}\ \bibnamefont
  {Hartsock}}\ and\ \bibinfo {author} {\bibfnamefont {W.~James}\ \bibnamefont
  {Nelson}},\ }\bibfield  {title} {\enquote {\bibinfo {title} {Adherens and
  {Tight} {Junctions}: {Structure}, {Function} and {Connections} to the {Actin}
  {Cytoskeleton}},}\ }\href {\doibase 10.1016/j.bbamem.2007.07.012} {\bibfield
  {journal} {\bibinfo  {journal} {Biochimica et biophysica acta}\ }\textbf
  {\bibinfo {volume} {1778}},\ \bibinfo {pages} {660--669} (\bibinfo {year}
  {2008})}\BibitemShut {NoStop}%
\bibitem [{\citenamefont {Lodish}\ \emph {et~al.}(2016)\citenamefont {Lodish},
  \citenamefont {Berk}, \citenamefont {Kaiser}, \citenamefont {Krieger},
  \citenamefont {Bretscher}, \citenamefont {Ploegh}, \citenamefont {Amon},\
  and\ \citenamefont {Martin}}]{lodish2008molecular}%
  \BibitemOpen
  \bibfield  {author} {\bibinfo {author} {\bibfnamefont {Harvey}\ \bibnamefont
  {Lodish}}, \bibinfo {author} {\bibfnamefont {Arnold}\ \bibnamefont {Berk}},
  \bibinfo {author} {\bibfnamefont {Chris~A.}\ \bibnamefont {Kaiser}}, \bibinfo
  {author} {\bibfnamefont {Monty}\ \bibnamefont {Krieger}}, \bibinfo {author}
  {\bibfnamefont {Anthony}\ \bibnamefont {Bretscher}}, \bibinfo {author}
  {\bibfnamefont {Hidde}\ \bibnamefont {Ploegh}}, \bibinfo {author}
  {\bibfnamefont {Angelika}\ \bibnamefont {Amon}}, \ and\ \bibinfo {author}
  {\bibfnamefont {Kelsey~C.}\ \bibnamefont {Martin}},\ }\href@noop {}
  {{\selectlanguage {english}\emph {\bibinfo {title} {Molecular {Cell}
  {Biology}}}}},\ \bibinfo {edition} {eighth edition}\ ed.\ (\bibinfo
  {publisher} {W. H. Freeman},\ \bibinfo {address} {New York},\ \bibinfo {year}
  {2016})\BibitemShut {NoStop}%
\bibitem [{\citenamefont {Halbleib}\ and\ \citenamefont
  {Nelson}(2006)}]{Halbleib06}%
  \BibitemOpen
  \bibfield  {author} {\bibinfo {author} {\bibfnamefont {Jennifer~M}\
  \bibnamefont {Halbleib}}\ and\ \bibinfo {author} {\bibfnamefont {W~James}\
  \bibnamefont {Nelson}},\ }\bibfield  {title} {\enquote {\bibinfo {title}
  {Cadherins in development: cell adhesion, sorting, and tissue
  morphogenesis},}\ }\href@noop {} {\bibfield  {journal} {\bibinfo  {journal}
  {Genes \& development}\ }\textbf {\bibinfo {volume} {20}},\ \bibinfo {pages}
  {3199--3214} (\bibinfo {year} {2006})}\BibitemShut {NoStop}%
\bibitem [{\citenamefont {Franke}\ \emph {et~al.}(2005)\citenamefont {Franke},
  \citenamefont {Montague},\ and\ \citenamefont {Kiehart}}]{Frank05}%
  \BibitemOpen
  \bibfield  {author} {\bibinfo {author} {\bibfnamefont {Josef~D}\ \bibnamefont
  {Franke}}, \bibinfo {author} {\bibfnamefont {Ruth~A}\ \bibnamefont
  {Montague}}, \ and\ \bibinfo {author} {\bibfnamefont {Daniel~P}\ \bibnamefont
  {Kiehart}},\ }\bibfield  {title} {\enquote {\bibinfo {title} {Nonmuscle
  myosin ii generates forces that transmit tension and drive contraction in
  multiple tissues during dorsal closure},}\ }\href@noop {} {\bibfield
  {journal} {\bibinfo  {journal} {Current Biology}\ }\textbf {\bibinfo {volume}
  {15}},\ \bibinfo {pages} {2208--2221} (\bibinfo {year} {2005})}\BibitemShut
  {NoStop}%
\bibitem [{\citenamefont {Salbreux}\ \emph {et~al.}(2012)\citenamefont
  {Salbreux}, \citenamefont {Charras},\ and\ \citenamefont
  {Paluch}}]{Paluch2012}%
  \BibitemOpen
  \bibfield  {author} {\bibinfo {author} {\bibfnamefont {Guillaume}\
  \bibnamefont {Salbreux}}, \bibinfo {author} {\bibfnamefont {Guillaume}\
  \bibnamefont {Charras}}, \ and\ \bibinfo {author} {\bibfnamefont {Ewa}\
  \bibnamefont {Paluch}},\ }\bibfield  {title} {\enquote {\bibinfo {title}
  {Actin cortex mechanics and cellular morphogenesis},}\ }\href@noop {}
  {\bibfield  {journal} {\bibinfo  {journal} {Trends in cell biology}\ }\textbf
  {\bibinfo {volume} {22}},\ \bibinfo {pages} {536--545} (\bibinfo {year}
  {2012})}\BibitemShut {NoStop}%
\bibitem [{\citenamefont {Stewart}\ \emph {et~al.}(2011)\citenamefont
  {Stewart}, \citenamefont {Helenius}, \citenamefont {Toyoda}, \citenamefont
  {Ramanathan}, \citenamefont {Muller},\ and\ \citenamefont
  {Hyman}}]{Hyman2011}%
  \BibitemOpen
  \bibfield  {author} {\bibinfo {author} {\bibfnamefont {Martin~P}\
  \bibnamefont {Stewart}}, \bibinfo {author} {\bibfnamefont {Jonne}\
  \bibnamefont {Helenius}}, \bibinfo {author} {\bibfnamefont {Yusuke}\
  \bibnamefont {Toyoda}}, \bibinfo {author} {\bibfnamefont {Subramanian~P}\
  \bibnamefont {Ramanathan}}, \bibinfo {author} {\bibfnamefont {Daniel~J}\
  \bibnamefont {Muller}}, \ and\ \bibinfo {author} {\bibfnamefont {Anthony~A}\
  \bibnamefont {Hyman}},\ }\bibfield  {title} {\enquote {\bibinfo {title}
  {Hydrostatic pressure and the actomyosin cortex drive mitotic cell
  rounding},}\ }\href@noop {} {\bibfield  {journal} {\bibinfo  {journal}
  {Nature}\ }\textbf {\bibinfo {volume} {469}},\ \bibinfo {pages} {226}
  (\bibinfo {year} {2011})}\BibitemShut {NoStop}%
\bibitem [{\citenamefont {Honda}\ \emph {et~al.}(1984)\citenamefont {Honda},
  \citenamefont {Yamanaka},\ and\ \citenamefont {Dan-Sohkawa}}]{Honda83}%
  \BibitemOpen
  \bibfield  {author} {\bibinfo {author} {\bibfnamefont {H}~\bibnamefont
  {Honda}}, \bibinfo {author} {\bibfnamefont {H}~\bibnamefont {Yamanaka}}, \
  and\ \bibinfo {author} {\bibfnamefont {M}~\bibnamefont {Dan-Sohkawa}},\
  }\bibfield  {title} {\enquote {\bibinfo {title} {A computer simulation of
  geometrical configurations during cell division},}\ }\href@noop {} {\bibfield
   {journal} {\bibinfo  {journal} {Journal of theoretical biology}\ }\textbf
  {\bibinfo {volume} {106}},\ \bibinfo {pages} {423--435} (\bibinfo {year}
  {1984})}\BibitemShut {NoStop}%
\bibitem [{\citenamefont {Hufnagel}\ \emph {et~al.}(2007)\citenamefont
  {Hufnagel}, \citenamefont {Teleman}, \citenamefont {Rouault}, \citenamefont
  {Cohen},\ and\ \citenamefont {Shraiman}}]{Huf07}%
  \BibitemOpen
  \bibfield  {author} {\bibinfo {author} {\bibfnamefont {Lars}\ \bibnamefont
  {Hufnagel}}, \bibinfo {author} {\bibfnamefont {Aurelio~A}\ \bibnamefont
  {Teleman}}, \bibinfo {author} {\bibfnamefont {Herv{\'e}}\ \bibnamefont
  {Rouault}}, \bibinfo {author} {\bibfnamefont {Stephen~M}\ \bibnamefont
  {Cohen}}, \ and\ \bibinfo {author} {\bibfnamefont {Boris~I}\ \bibnamefont
  {Shraiman}},\ }\bibfield  {title} {\enquote {\bibinfo {title} {On the
  mechanism of wing size determination in fly development},}\ }\href@noop {}
  {\bibfield  {journal} {\bibinfo  {journal} {Proceedings of the National
  Academy of Sciences}\ }\textbf {\bibinfo {volume} {104}},\ \bibinfo {pages}
  {3835--3840} (\bibinfo {year} {2007})}\BibitemShut {NoStop}%
\bibitem [{\citenamefont {Noll}\ \emph {et~al.}(2017)\citenamefont {Noll},
  \citenamefont {Mani}, \citenamefont {Heemskerk}, \citenamefont {Streichan},\
  and\ \citenamefont {Shraiman}}]{Noll17}%
  \BibitemOpen
  \bibfield  {author} {\bibinfo {author} {\bibfnamefont {Nicholas}\
  \bibnamefont {Noll}}, \bibinfo {author} {\bibfnamefont {Madhav}\ \bibnamefont
  {Mani}}, \bibinfo {author} {\bibfnamefont {Idse}\ \bibnamefont {Heemskerk}},
  \bibinfo {author} {\bibfnamefont {Sebastian~J}\ \bibnamefont {Streichan}}, \
  and\ \bibinfo {author} {\bibfnamefont {Boris~I}\ \bibnamefont {Shraiman}},\
  }\bibfield  {title} {\enquote {\bibinfo {title} {Active tension network model
  suggests an exotic mechanical state realized in epithelial tissues},}\
  }\href@noop {} {\bibfield  {journal} {\bibinfo  {journal} {Nature physics}\
  }\textbf {\bibinfo {volume} {13}},\ \bibinfo {pages} {1221} (\bibinfo {year}
  {2017})}\BibitemShut {NoStop}%
\bibitem [{\citenamefont {Haase}\ and\ \citenamefont
  {Pelling}(2015)}]{AFM_review}%
  \BibitemOpen
  \bibfield  {author} {\bibinfo {author} {\bibfnamefont {Kristina}\
  \bibnamefont {Haase}}\ and\ \bibinfo {author} {\bibfnamefont {Andrew~E.}\
  \bibnamefont {Pelling}},\ }\bibfield  {title} {{\selectlanguage
  {english}\enquote {\bibinfo {title} {Investigating cell mechanics with atomic
  force microscopy},}\ }}\href {\doibase 10.1098/rsif.2014.0970} {\bibfield
  {journal} {\bibinfo  {journal} {Journal of The Royal Society Interface}\
  }\textbf {\bibinfo {volume} {12}},\ \bibinfo {pages} {20140970} (\bibinfo
  {year} {2015})}\BibitemShut {NoStop}%
\bibitem [{\citenamefont {Bambardekar}\ \emph {et~al.}(2015)\citenamefont
  {Bambardekar}, \citenamefont {Cl{\'e}ment}, \citenamefont {Blanc},
  \citenamefont {Chard{\`e}s},\ and\ \citenamefont {Lenne}}]{Bam09}%
  \BibitemOpen
  \bibfield  {author} {\bibinfo {author} {\bibfnamefont {Kapil}\ \bibnamefont
  {Bambardekar}}, \bibinfo {author} {\bibfnamefont {Rapha{\"e}l}\ \bibnamefont
  {Cl{\'e}ment}}, \bibinfo {author} {\bibfnamefont {Olivier}\ \bibnamefont
  {Blanc}}, \bibinfo {author} {\bibfnamefont {Claire}\ \bibnamefont
  {Chard{\`e}s}}, \ and\ \bibinfo {author} {\bibfnamefont
  {Pierre-Fran{\c{c}}ois}\ \bibnamefont {Lenne}},\ }\bibfield  {title}
  {\enquote {\bibinfo {title} {Direct laser manipulation reveals the mechanics
  of cell contacts in vivo},}\ }\href@noop {} {\bibfield  {journal} {\bibinfo
  {journal} {Proceedings of the National Academy of Sciences}\ }\textbf
  {\bibinfo {volume} {112}},\ \bibinfo {pages} {1416--1421} (\bibinfo {year}
  {2015})}\BibitemShut {NoStop}%
\bibitem [{\citenamefont {Bonnet}\ \emph {et~al.}(2012)\citenamefont {Bonnet},
  \citenamefont {Marcq}, \citenamefont {Bosveld}, \citenamefont {Fetler},
  \citenamefont {Bella{\"\i}che},\ and\ \citenamefont {Graner}}]{Bonnet12}%
  \BibitemOpen
  \bibfield  {author} {\bibinfo {author} {\bibfnamefont {Isabelle}\
  \bibnamefont {Bonnet}}, \bibinfo {author} {\bibfnamefont {Philippe}\
  \bibnamefont {Marcq}}, \bibinfo {author} {\bibfnamefont {Floris}\
  \bibnamefont {Bosveld}}, \bibinfo {author} {\bibfnamefont {Luc}\ \bibnamefont
  {Fetler}}, \bibinfo {author} {\bibfnamefont {Yohanns}\ \bibnamefont
  {Bella{\"\i}che}}, \ and\ \bibinfo {author} {\bibfnamefont {Fran{\c{c}}ois}\
  \bibnamefont {Graner}},\ }\bibfield  {title} {\enquote {\bibinfo {title}
  {Mechanical state, material properties and continuous description of an
  epithelial tissue},}\ }\href@noop {} {\bibfield  {journal} {\bibinfo
  {journal} {Journal of The Royal Society Interface}\ }\textbf {\bibinfo
  {volume} {9}},\ \bibinfo {pages} {2614--2623} (\bibinfo {year}
  {2012})}\BibitemShut {NoStop}%
\bibitem [{\citenamefont {Cost}\ \emph {et~al.}(2015)\citenamefont {Cost},
  \citenamefont {Ringer}, \citenamefont {Chrostek-Grashoff},\ and\
  \citenamefont {Grashoff}}]{FRET_Grashoff}%
  \BibitemOpen
  \bibfield  {author} {\bibinfo {author} {\bibfnamefont {Anna-Lena}\
  \bibnamefont {Cost}}, \bibinfo {author} {\bibfnamefont {Pia}\ \bibnamefont
  {Ringer}}, \bibinfo {author} {\bibfnamefont {Anna}\ \bibnamefont
  {Chrostek-Grashoff}}, \ and\ \bibinfo {author} {\bibfnamefont {Carsten}\
  \bibnamefont {Grashoff}},\ }\bibfield  {title} {\enquote {\bibinfo {title}
  {How to {Measure} {Molecular} {Forces} in {Cells}: {A} {Guide} to
  {Evaluating} {Genetically}-{Encoded} {FRET}-{Based} {Tension} {Sensors}},}\
  }\href {\doibase 10.1007/s12195-014-0368-1} {\bibfield  {journal} {\bibinfo
  {journal} {Cellular and Molecular Bioengineering}\ }\textbf {\bibinfo
  {volume} {8}},\ \bibinfo {pages} {96--105} (\bibinfo {year}
  {2015})}\BibitemShut {NoStop}%
\bibitem [{\citenamefont {Campàs}\ \emph {et~al.}(2014)\citenamefont
  {Campàs}, \citenamefont {Mammoto}, \citenamefont {Hasso}, \citenamefont
  {Sperling}, \citenamefont {O’Connell}, \citenamefont {Bischof},
  \citenamefont {Maas}, \citenamefont {Weitz}, \citenamefont {Mahadevan},\ and\
  \citenamefont {Ingber}}]{Campas14}%
  \BibitemOpen
  \bibfield  {author} {\bibinfo {author} {\bibfnamefont {Otger}\ \bibnamefont
  {Campàs}}, \bibinfo {author} {\bibfnamefont {Tadanori}\ \bibnamefont
  {Mammoto}}, \bibinfo {author} {\bibfnamefont {Sean}\ \bibnamefont {Hasso}},
  \bibinfo {author} {\bibfnamefont {Ralph~A}\ \bibnamefont {Sperling}},
  \bibinfo {author} {\bibfnamefont {Daniel}\ \bibnamefont {O’Connell}},
  \bibinfo {author} {\bibfnamefont {Ashley~G}\ \bibnamefont {Bischof}},
  \bibinfo {author} {\bibfnamefont {Richard}\ \bibnamefont {Maas}}, \bibinfo
  {author} {\bibfnamefont {David~A}\ \bibnamefont {Weitz}}, \bibinfo {author}
  {\bibfnamefont {Lakshminarayanan}\ \bibnamefont {Mahadevan}}, \ and\ \bibinfo
  {author} {\bibfnamefont {Donald~E}\ \bibnamefont {Ingber}},\ }\bibfield
  {title} {\enquote {\bibinfo {title} {Quantifying cell-generated mechanical
  forces within living embryonic tissues},}\ }\href {\doibase
  10.1038/nmeth.2761} {\bibfield  {journal} {\bibinfo  {journal} {Nature
  methods}\ }\textbf {\bibinfo {volume} {11}},\ \bibinfo {pages} {183--189}
  (\bibinfo {year} {2014})}\BibitemShut {NoStop}%
\bibitem [{\citenamefont {Doubrovinski}\ \emph {et~al.}(2017)\citenamefont
  {Doubrovinski}, \citenamefont {Swan}, \citenamefont {Polyakov},\ and\
  \citenamefont {Wieschaus}}]{Doub17}%
  \BibitemOpen
  \bibfield  {author} {\bibinfo {author} {\bibfnamefont {Konstantin}\
  \bibnamefont {Doubrovinski}}, \bibinfo {author} {\bibfnamefont {Michael}\
  \bibnamefont {Swan}}, \bibinfo {author} {\bibfnamefont {Oleg}\ \bibnamefont
  {Polyakov}}, \ and\ \bibinfo {author} {\bibfnamefont {{Eric F.}}\
  \bibnamefont {Wieschaus}},\ }\bibfield  {title} {{\selectlanguage
  {english}\enquote {\bibinfo {title} {Measurement of cortical elasticity in
  drosophila melanogaster embryos using ferrofluids},}\ }}\href {\doibase
  10.1073/pnas.1616659114} {\bibfield  {journal} {\bibinfo  {journal}
  {Proceedings of the National Academy of Sciences of the United States of
  America}\ }\textbf {\bibinfo {volume} {114}},\ \bibinfo {pages} {1051--1056}
  (\bibinfo {year} {2017})}\BibitemShut {NoStop}%
\bibitem [{\citenamefont {Brodland}\ \emph {et~al.}(2010)\citenamefont
  {Brodland}, \citenamefont {Conte}, \citenamefont {Cranston}, \citenamefont
  {Veldhuis}, \citenamefont {Narasimhan}, \citenamefont {Hutson}, \citenamefont
  {Jacinto}, \citenamefont {Ulrich}, \citenamefont {Baum},\ and\ \citenamefont
  {Miodownik}}]{Brodland10}%
  \BibitemOpen
  \bibfield  {author} {\bibinfo {author} {\bibfnamefont {G~Wayne}\ \bibnamefont
  {Brodland}}, \bibinfo {author} {\bibfnamefont {Vito}\ \bibnamefont {Conte}},
  \bibinfo {author} {\bibfnamefont {P~Graham}\ \bibnamefont {Cranston}},
  \bibinfo {author} {\bibfnamefont {Jim}\ \bibnamefont {Veldhuis}}, \bibinfo
  {author} {\bibfnamefont {Sriram}\ \bibnamefont {Narasimhan}}, \bibinfo
  {author} {\bibfnamefont {M~Shane}\ \bibnamefont {Hutson}}, \bibinfo {author}
  {\bibfnamefont {Antonio}\ \bibnamefont {Jacinto}}, \bibinfo {author}
  {\bibfnamefont {Florian}\ \bibnamefont {Ulrich}}, \bibinfo {author}
  {\bibfnamefont {Buzz}\ \bibnamefont {Baum}}, \ and\ \bibinfo {author}
  {\bibfnamefont {Mark}\ \bibnamefont {Miodownik}},\ }\bibfield  {title}
  {\enquote {\bibinfo {title} {Video force microscopy reveals the mechanics of
  ventral furrow invagination in drosophila},}\ }\href@noop {} {\bibfield
  {journal} {\bibinfo  {journal} {Proceedings of the National Academy of
  Sciences}\ }\textbf {\bibinfo {volume} {107}},\ \bibinfo {pages}
  {22111--22116} (\bibinfo {year} {2010})}\BibitemShut {NoStop}%
\bibitem [{\citenamefont {Chiou}\ \emph {et~al.}(2012)\citenamefont {Chiou},
  \citenamefont {Hufnagel},\ and\ \citenamefont {Shraiman}}]{Chiou12}%
  \BibitemOpen
  \bibfield  {author} {\bibinfo {author} {\bibfnamefont {Kevin~K}\ \bibnamefont
  {Chiou}}, \bibinfo {author} {\bibfnamefont {Lars}\ \bibnamefont {Hufnagel}},
  \ and\ \bibinfo {author} {\bibfnamefont {Boris~I}\ \bibnamefont {Shraiman}},\
  }\bibfield  {title} {\enquote {\bibinfo {title} {Mechanical stress inference
  for two dimensional cell arrays},}\ }\href@noop {} {\bibfield  {journal}
  {\bibinfo  {journal} {PLoS computational biology}\ }\textbf {\bibinfo
  {volume} {8}},\ \bibinfo {pages} {e1002512} (\bibinfo {year}
  {2012})}\BibitemShut {NoStop}%
\bibitem [{\citenamefont {Ishihara}\ and\ \citenamefont
  {Sugimura}(2012)}]{Ishihara12}%
  \BibitemOpen
  \bibfield  {author} {\bibinfo {author} {\bibfnamefont {Shuji}\ \bibnamefont
  {Ishihara}}\ and\ \bibinfo {author} {\bibfnamefont {Kaoru}\ \bibnamefont
  {Sugimura}},\ }\bibfield  {title} {\enquote {\bibinfo {title} {Bayesian
  inference of force dynamics during morphogenesis},}\ }\href@noop {}
  {\bibfield  {journal} {\bibinfo  {journal} {Journal of theoretical biology}\
  }\textbf {\bibinfo {volume} {313}},\ \bibinfo {pages} {201--211} (\bibinfo
  {year} {2012})}\BibitemShut {NoStop}%
\bibitem [{\citenamefont {Brodland}\ \emph {et~al.}(2014)\citenamefont
  {Brodland}, \citenamefont {Veldhuis}, \citenamefont {Kim}, \citenamefont
  {Perrone}, \citenamefont {Mashburn},\ and\ \citenamefont {Hutson}}]{CellFit}%
  \BibitemOpen
  \bibfield  {author} {\bibinfo {author} {\bibfnamefont {G.~W.}\ \bibnamefont
  {Brodland}}, \bibinfo {author} {\bibfnamefont {J.}~\bibnamefont {Veldhuis}},
  \bibinfo {author} {\bibfnamefont {S.}~\bibnamefont {Kim}}, \bibinfo {author}
  {\bibfnamefont {M.}~\bibnamefont {Perrone}}, \bibinfo {author} {\bibfnamefont
  {D.}~\bibnamefont {Mashburn}}, \ and\ \bibinfo {author} {\bibfnamefont
  {M.}~\bibnamefont {Hutson}},\ }\bibfield  {title} {\enquote {\bibinfo {title}
  {Cellfit: A cellular force-inference toolkit using curvilinear cell
  boundaries},}\ }\href@noop {} {\bibfield  {journal} {\bibinfo  {journal}
  {PLoS One}\ }\textbf {\bibinfo {volume} {9}} (\bibinfo {year}
  {2014})}\BibitemShut {NoStop}%
\bibitem [{\citenamefont {Streichan}\ \emph {et~al.}(2018)\citenamefont
  {Streichan}, \citenamefont {Lefebvre}, \citenamefont {Noll}, \citenamefont
  {Wieschaus},\ and\ \citenamefont {Shraiman}}]{Streichan17}%
  \BibitemOpen
  \bibfield  {author} {\bibinfo {author} {\bibfnamefont {Sebastian~J}\
  \bibnamefont {Streichan}}, \bibinfo {author} {\bibfnamefont {Matthew~F}\
  \bibnamefont {Lefebvre}}, \bibinfo {author} {\bibfnamefont {Nicholas}\
  \bibnamefont {Noll}}, \bibinfo {author} {\bibfnamefont {Eric~F}\ \bibnamefont
  {Wieschaus}}, \ and\ \bibinfo {author} {\bibfnamefont {Boris~I}\ \bibnamefont
  {Shraiman}},\ }\bibfield  {title} {\enquote {\bibinfo {title} {Global
  morphogenetic flow is accurately predicted by the spatial distribution of
  myosin motors},}\ }\href {\doibase 10.7554/eLife.27454} {\bibfield  {journal}
  {\bibinfo  {journal} {eLife}\ }\textbf {\bibinfo {volume} {7}},\ \bibinfo
  {pages} {e27454} (\bibinfo {year} {2018})}\BibitemShut {NoStop}%
\bibitem [{\citenamefont {He}\ \emph {et~al.}(2014)\citenamefont {He},
  \citenamefont {Doubrovinski}, \citenamefont {Polyakov},\ and\ \citenamefont
  {Wieschaus}}]{Bing14}%
  \BibitemOpen
  \bibfield  {author} {\bibinfo {author} {\bibfnamefont {Bing}\ \bibnamefont
  {He}}, \bibinfo {author} {\bibfnamefont {Konstantin}\ \bibnamefont
  {Doubrovinski}}, \bibinfo {author} {\bibfnamefont {Oleg}\ \bibnamefont
  {Polyakov}}, \ and\ \bibinfo {author} {\bibfnamefont {Eric}\ \bibnamefont
  {Wieschaus}},\ }\bibfield  {title} {\enquote {\bibinfo {title} {Apical
  constriction drives tissue-scale hydrodynamic flow to mediate cell
  elongation},}\ }\href {\doibase 10.1038/nature13070} {\bibfield  {journal}
  {\bibinfo  {journal} {Nature}\ }\textbf {\bibinfo {volume} {508}},\ \bibinfo
  {pages} {392--396} (\bibinfo {year} {2014})}\BibitemShut {NoStop}%
\bibitem [{\citenamefont {Collinet}\ \emph {et~al.}(2015)\citenamefont
  {Collinet}, \citenamefont {Rauzi}, \citenamefont {Lenne},\ and\ \citenamefont
  {Lecuit}}]{Collinet15}%
  \BibitemOpen
  \bibfield  {author} {\bibinfo {author} {\bibfnamefont {Claudio}\ \bibnamefont
  {Collinet}}, \bibinfo {author} {\bibfnamefont {Matteo}\ \bibnamefont
  {Rauzi}}, \bibinfo {author} {\bibfnamefont {Pierre-Fran{\c{c}}ois}\
  \bibnamefont {Lenne}}, \ and\ \bibinfo {author} {\bibfnamefont {Thomas}\
  \bibnamefont {Lecuit}},\ }\bibfield  {title} {\enquote {\bibinfo {title}
  {Local and tissue-scale forces drive oriented junction growth during tissue
  extension},}\ }\href@noop {} {\bibfield  {journal} {\bibinfo  {journal}
  {Nature cell biology}\ }\textbf {\bibinfo {volume} {17}},\ \bibinfo {pages}
  {1247} (\bibinfo {year} {2015})}\BibitemShut {NoStop}%
\bibitem [{\citenamefont {{L.D. Landau and E.M. Lifshitz}}()}]{landau}%
  \BibitemOpen
  \bibfield  {author} {\bibinfo {author} {\bibnamefont {{L.D. Landau and E.M.
  Lifshitz}}},\ }\href {http://archive.org/details/FluidMechanics} {\emph
  {\bibinfo {title} {Fluid {Mechanics}}}}\BibitemShut {NoStop}%
\bibitem [{\citenamefont {Zallen}\ and\ \citenamefont
  {Blankenship}(2008)}]{Zallen08}%
  \BibitemOpen
  \bibfield  {author} {\bibinfo {author} {\bibfnamefont {Jennifer~A.}\
  \bibnamefont {Zallen}}\ and\ \bibinfo {author} {\bibfnamefont {J.~Todd}\
  \bibnamefont {Blankenship}},\ }\bibfield  {title} {\enquote {\bibinfo {title}
  {Multicellular dynamics during epithelial elongation},}\ }\href {\doibase
  10.1016/j.semcdb.2008.01.005} {\bibfield  {journal} {\bibinfo  {journal}
  {Seminars in cell \& developmental biology}\ }\textbf {\bibinfo {volume}
  {19}},\ \bibinfo {pages} {263--270} (\bibinfo {year} {2008})}\BibitemShut
  {NoStop}%
\bibitem [{\citenamefont {Rauzi}\ \emph {et~al.}(2010)\citenamefont {Rauzi},
  \citenamefont {Lenne},\ and\ \citenamefont {Lecuit}}]{Rauz10}%
  \BibitemOpen
  \bibfield  {author} {\bibinfo {author} {\bibfnamefont {Matteo}\ \bibnamefont
  {Rauzi}}, \bibinfo {author} {\bibfnamefont {Pierre-Fran{\c{c}}ois}\
  \bibnamefont {Lenne}}, \ and\ \bibinfo {author} {\bibfnamefont {Thomas}\
  \bibnamefont {Lecuit}},\ }\bibfield  {title} {\enquote {\bibinfo {title}
  {Planar polarized actomyosin contractile flows control epithelial junction
  remodelling},}\ }\href@noop {} {\bibfield  {journal} {\bibinfo  {journal}
  {Nature}\ }\textbf {\bibinfo {volume} {468}},\ \bibinfo {pages} {1110}
  (\bibinfo {year} {2010})}\BibitemShut {NoStop}%
\bibitem [{\citenamefont {Heemskerk}\ and\ \citenamefont
  {Streichan}(2015)}]{Heemskerk15}%
  \BibitemOpen
  \bibfield  {author} {\bibinfo {author} {\bibfnamefont {Idse}\ \bibnamefont
  {Heemskerk}}\ and\ \bibinfo {author} {\bibfnamefont {Sebastian~J}\
  \bibnamefont {Streichan}},\ }\bibfield  {title} {\enquote {\bibinfo {title}
  {Tissue cartography: compressing bio-image data by dimensional reduction},}\
  }\href@noop {} {\bibfield  {journal} {\bibinfo  {journal} {Nature methods}\
  }\textbf {\bibinfo {volume} {12}},\ \bibinfo {pages} {1139} (\bibinfo {year}
  {2015})}\BibitemShut {NoStop}%
\bibitem [{\citenamefont {Foe}(1989)}]{Foe89}%
  \BibitemOpen
  \bibfield  {author} {\bibinfo {author} {\bibfnamefont {V.~E.}\ \bibnamefont
  {Foe}},\ }\bibfield  {title} {{\selectlanguage {english}\enquote {\bibinfo
  {title} {Mitotic domains reveal early commitment of cells in {Drosophila}
  embryos},}\ }}\href {http://dev.biologists.org/content/107/1/1} {\bibfield
  {journal} {\bibinfo  {journal} {Development}\ }\textbf {\bibinfo {volume}
  {107}},\ \bibinfo {pages} {1--22} (\bibinfo {year} {1989})}\BibitemShut
  {NoStop}%
\bibitem [{\citenamefont {da~Silva}\ and\ \citenamefont
  {Vincent}(2007)}]{Morais07}%
  \BibitemOpen
  \bibfield  {author} {\bibinfo {author} {\bibfnamefont {Sara~Morais}\
  \bibnamefont {da~Silva}}\ and\ \bibinfo {author} {\bibfnamefont {Jean-Paul}\
  \bibnamefont {Vincent}},\ }\bibfield  {title} {\enquote {\bibinfo {title}
  {Oriented cell divisions in the extending germband of drosophila},}\
  }\href@noop {} {\bibfield  {journal} {\bibinfo  {journal} {Development}\
  }\textbf {\bibinfo {volume} {134}},\ \bibinfo {pages} {3049--3054} (\bibinfo
  {year} {2007})}\BibitemShut {NoStop}%
\bibitem [{\citenamefont {Louveaux}\ \emph {et~al.}(2016)\citenamefont
  {Louveaux}, \citenamefont {Julien}, \citenamefont {Mirabet}, \citenamefont
  {Boudaoud},\ and\ \citenamefont {Hamant}}]{Louveaux16}%
  \BibitemOpen
  \bibfield  {author} {\bibinfo {author} {\bibfnamefont {Marion}\ \bibnamefont
  {Louveaux}}, \bibinfo {author} {\bibfnamefont {Jean-Daniel}\ \bibnamefont
  {Julien}}, \bibinfo {author} {\bibfnamefont {Vincent}\ \bibnamefont
  {Mirabet}}, \bibinfo {author} {\bibfnamefont {Arezki}\ \bibnamefont
  {Boudaoud}}, \ and\ \bibinfo {author} {\bibfnamefont {Olivier}\ \bibnamefont
  {Hamant}},\ }\bibfield  {title} {\enquote {\bibinfo {title} {Cell division
  plane orientation based on tensile stress in arabidopsis thaliana},}\
  }\href@noop {} {\bibfield  {journal} {\bibinfo  {journal} {Proceedings of the
  National Academy of Sciences}\ }\textbf {\bibinfo {volume} {113}},\ \bibinfo
  {pages} {E4294--E4303} (\bibinfo {year} {2016})}\BibitemShut {NoStop}%
\bibitem [{\citenamefont {Nestor-Bergmann}\ \emph {et~al.}(2014)\citenamefont
  {Nestor-Bergmann}, \citenamefont {Goddard},\ and\ \citenamefont
  {Woolner}}]{Nestor14}%
  \BibitemOpen
  \bibfield  {author} {\bibinfo {author} {\bibfnamefont {Alexander}\
  \bibnamefont {Nestor-Bergmann}}, \bibinfo {author} {\bibfnamefont {Georgina}\
  \bibnamefont {Goddard}}, \ and\ \bibinfo {author} {\bibfnamefont {Sarah}\
  \bibnamefont {Woolner}},\ }\bibfield  {title} {\enquote {\bibinfo {title}
  {Force and the spindle: mechanical cues in mitotic spindle orientation},}\
  }in\ \href@noop {} {\emph {\bibinfo {booktitle} {Seminars in cell \&
  developmental biology}}},\ Vol.~\bibinfo {volume} {34}\ (\bibinfo
  {organization} {Elsevier},\ \bibinfo {year} {2014})\ pp.\ \bibinfo {pages}
  {133--139}\BibitemShut {NoStop}%
\bibitem [{\citenamefont {Krzic}\ \emph {et~al.}(2012)\citenamefont {Krzic},
  \citenamefont {Gunther}, \citenamefont {Saunders}, \citenamefont
  {Streichan},\ and\ \citenamefont {Hufnagel}}]{Krzic2012}%
  \BibitemOpen
  \bibfield  {author} {\bibinfo {author} {\bibfnamefont {Uros}\ \bibnamefont
  {Krzic}}, \bibinfo {author} {\bibfnamefont {Stefan}\ \bibnamefont {Gunther}},
  \bibinfo {author} {\bibfnamefont {Timothy~E}\ \bibnamefont {Saunders}},
  \bibinfo {author} {\bibfnamefont {Sebastian~J}\ \bibnamefont {Streichan}}, \
  and\ \bibinfo {author} {\bibfnamefont {Lars}\ \bibnamefont {Hufnagel}},\
  }\bibfield  {title} {{\selectlanguage {english}\enquote {\bibinfo {title}
  {Multiview light-sheet microscope for rapid in toto imaging},}\ }}\href
  {\doibase 10.1038/nmeth.2064} {\bibfield  {journal} {\bibinfo  {journal}
  {Nature Methods}\ }\textbf {\bibinfo {volume} {9}},\ \bibinfo {pages}
  {730--733} (\bibinfo {year} {2012})}\BibitemShut {NoStop}%
\bibitem [{\citenamefont {Edelstein}\ \emph {et~al.}(2014)\citenamefont
  {Edelstein}, \citenamefont {Tsuchida}, \citenamefont {Amodaj}, \citenamefont
  {Pinkard}, \citenamefont {Vale},\ and\ \citenamefont
  {Stuurman}}]{Edelstein2014}%
  \BibitemOpen
  \bibfield  {author} {\bibinfo {author} {\bibfnamefont {Arthur~D.}\
  \bibnamefont {Edelstein}}, \bibinfo {author} {\bibfnamefont {Mark~A.}\
  \bibnamefont {Tsuchida}}, \bibinfo {author} {\bibfnamefont {Nenad}\
  \bibnamefont {Amodaj}}, \bibinfo {author} {\bibfnamefont {Henry}\
  \bibnamefont {Pinkard}}, \bibinfo {author} {\bibfnamefont {Ronald~D.}\
  \bibnamefont {Vale}}, \ and\ \bibinfo {author} {\bibfnamefont {Nico}\
  \bibnamefont {Stuurman}},\ }\bibfield  {title} {\enquote {\bibinfo {title}
  {Advanced methods of microscope control using μ{Manager} software},}\ }\href
  {\doibase 10.14440/jbm.2014.36} {\bibfield  {journal} {\bibinfo  {journal}
  {Journal of biological methods}\ }\textbf {\bibinfo {volume} {1}} (\bibinfo
  {year} {2014}),\ 10.14440/jbm.2014.36}\BibitemShut {NoStop}%
\bibitem [{\citenamefont {Preibisch}\ \emph {et~al.}(2014)\citenamefont
  {Preibisch}, \citenamefont {Amat}, \citenamefont {Stamataki}, \citenamefont
  {Sarov}, \citenamefont {Singer}, \citenamefont {Myers},\ and\ \citenamefont
  {Tomancak}}]{Preibisch2014}%
  \BibitemOpen
  \bibfield  {author} {\bibinfo {author} {\bibfnamefont {Stephan}\ \bibnamefont
  {Preibisch}}, \bibinfo {author} {\bibfnamefont {Fernando}\ \bibnamefont
  {Amat}}, \bibinfo {author} {\bibfnamefont {Evangelia}\ \bibnamefont
  {Stamataki}}, \bibinfo {author} {\bibfnamefont {Mihail}\ \bibnamefont
  {Sarov}}, \bibinfo {author} {\bibfnamefont {Robert~H}\ \bibnamefont
  {Singer}}, \bibinfo {author} {\bibfnamefont {Eugene}\ \bibnamefont {Myers}},
  \ and\ \bibinfo {author} {\bibfnamefont {Pavel}\ \bibnamefont {Tomancak}},\
  }\bibfield  {title} {\enquote {\bibinfo {title} {Efficient {Bayesian}-based
  multiview deconvolution},}\ }\href {\doibase 10.1038/nmeth.2929} {\bibfield
  {journal} {\bibinfo  {journal} {Nature methods}\ }\textbf {\bibinfo {volume}
  {11}},\ \bibinfo {pages} {645--648} (\bibinfo {year} {2014})}\BibitemShut
  {NoStop}%
\bibitem [{\citenamefont {Walker}(2018)}]{Walker2018}%
  \BibitemOpen
  \bibfield  {author} {\bibinfo {author} {\bibfnamefont {Shawn~W.}\
  \bibnamefont {Walker}},\ }\bibfield  {title} {{\selectlanguage
  {english}\enquote {\bibinfo {title} {{FELICITY}: {A} {Matlab}/{C}++ {Toolbox}
  for {Developing} {Finite} {Element} {Methods} and {Simulation} {Modeling}},}\
  }}\href {\doibase 10.1137/17M1128745} {\bibfield  {journal} {\bibinfo
  {journal} {SIAM Journal on Scientific Computing}\ }\textbf {\bibinfo {volume}
  {40}},\ \bibinfo {pages} {C234--C257} (\bibinfo {year} {2018})}\BibitemShut
  {NoStop}%
\end{thebibliography}%

\end{document}